\documentclass[11pt,reqno]{amsart}

\usepackage{amsfonts}
\usepackage{amsmath}
\usepackage{amssymb}
\usepackage{latexsym}
\usepackage{graphicx}
\usepackage{cite}

\numberwithin{equation}{section}

\def\deq{\simeq}

\newtheorem{thm}[equation]{Theorem}

\newtheorem{definition}[equation]{Definition}

\newcommand{\secref}[1]{section~\ref{#1}}

\numberwithin{equation}{section}

\renewcommand\a{\alpha}
\renewcommand\b{\beta}
\newcommand\g{\gamma}
\renewcommand\d{\delta}

\renewcommand\l{\lambda}

\newcommand\G{\Gamma}
\newcommand\f{\frac}

\newcommand\srel[2]{\begin{smallmatrix} {#1} \\ {#2} \end{smallmatrix}}

\newcommand{\Z}{{\mathbb{Z}}}

\newcommand{\C}{{\mathbb{C}}}
\newcommand{\A}{{\mathbb{A}}}
\newcommand{\Q}{{\mathbb{Q}}}

\renewcommand\Im{\mbox{Im~}}

\newcommand{\gobble}[1]{}
  \newcommand{\rangeref}[2]{%
    \ref{#1}--\afterassignment\gobble\fam 0\ref{#2}%
  }
\makeatletter
\def\imod#1{\allowbreak\mkern10mu({\operator@font mod}\,\,#1)}
\makeatother

\def\ZZ{{\mathbb Z}}

\def\IR{{\mathbb R}}
\def\RR{{\mathbb R}}

\def\IZ{{\mathbb Z}}

\def\cE{{\cal E}}

\def\cE{\mathcal{E}}
\def\cF{\mathcal{F}}
\def\cV{\mathcal{V}}
\def\cR{\mathcal{R}}

\newcommand{\be}{\begin{equation}}
\newcommand{\ee}{\end{equation}}
\newcommand{\bea}{\begin{eqnarray}}
\newcommand{\eea}{\end{eqnarray}}
\def\p{\partial }

\def\a{\alpha }
\def\g{\gamma }

\def\threeh{{\scriptstyle {3 \over 2}}}
\def\fiveh{{\scriptstyle {5 \over 2}}}

\def\bE{{\bf E}}

\def\hbE{{\hat{\bf E}}}

\def\calE{{\mathcal E}}

\def\calT{{\mathcal T}}

\def\bs{\backslash}

\def\calF{{\mathcal F}}

\def\calM{{\mathcal M}}
\def\calV{{\mathcal V}}

\def\nn{\nonumber}

\def\third{{\scriptstyle {1 \over 3}}}

\title[Eisenstein series and  string theory]{\bf Eisenstein series for higher-rank groups and string theory amplitudes}

\author[M.B. Green]{Michael B. Green}
\address{Michael B. Green\\
Department of Applied Mathematics and
Theoretical Physics\\
 Wilberforce Road, Cambridge CB3 0WA, UK}
\email{ M.B.Green@damtp.cam.ac.uk}
\author[S.D. Miller]{Stephen D. Miller}
\address{Stephen D Miller\\
Department of Mathematics\\
 Rutgers University, Piscataway, NJ 08854-8019, USA}
\email{miller@math.rutgers.edu}
\author[J.G. Russo]{Jorge G. Russo}
 \address{Jorge G Russo\\
 Instituci{\'o} Catalana de Recerca i Estudis Avan\c cats (ICREA)\\
 Departament ECM and Institut de Ciencies del Cosmos, \\
 University de Barcelona, Facultat de Fisica\\
 Av. Diagonal, 647, Barcelona 08028 Spain}
\email{jrusso@ub.edu}
\author[P. Vanhove]{Pierre Vanhove}
 \address{Pierre Vanhove\\
 Institut des Hautes Etudes Scientifiques\\
 Le Bois-Marie, 35 route de Chartres\\
 F-91440 Bures-sur-Yvette, France\hfill\break
Institut de Physique Th{\'e}orique,\\
CEA, IPhT, F-91191 Gif-sur-Yvette, France\\
CNRS, URA 2306, F-91191 Gif-sur-Yvette, France}
\email{pierre.vanhove@cea.fr}

\thanks{DAMTP-25-03-2010, IPHT-T-10/039, IHES/P/10/10, ICCUB-10-022}
\date{}

\begin{document}

 \begin{abstract}
 Scattering amplitudes of superstring theory are strongly constrained by the requirement that they be invariant  under dualities generated by discrete subgroups, $E_n(\IZ)$,  of  simply-laced Lie groups in the  $E_n$ series ($n\le 8$).
In particular, expanding the four-supergraviton amplitude at low energy gives a series of higher derivative corrections to Einstein's theory, with coefficients that are automorphic functions with a rich dependence on the moduli.
Boundary conditions supplied by string and supergravity perturbation theory, together with a chain of relations between successive   groups in the $E_n$ series,  constrain the  constant terms of these coefficients in three distinct parabolic subgroups.  Using this information we are able to determine the expressions for the first two higher derivative interactions (which are BPS-protected) in terms of specific Eisenstein series.  Further,  we determine key features of the coefficient of the third term in the low energy expansion of the four-supergraviton amplitude (which is also BPS-protected)   in the $E_8$ case.  This is an automorphic function that satisfies an inhomogeneous Laplace equation and has constant terms in certain parabolic subgroups that contain information about all the preceding terms.
\end{abstract}
\maketitle
\tableofcontents

\section{Introduction}

Superstring theory is highly constrained by dualities
combined  with  supersymmetry.     These  constraints  are  particularly
strong in  theories with maximal supersymmetry, which  can be obtained
by compactification of  ten-dimensional type II closed-string theories
on a $d$-torus, $\calT^d$, from $D=10$ dimensions to $D=10-d$, in which
case the theory is invariant under discrete subgroups $E_{d+1}(\ZZ)$
of the  real split  forms of the  Lie groups  $E_{d+1}$  defined in~\cite{Hull:1994ys,Chevalley}\footnote{The symbol $E_{d+1}$ will always
  refer  to the  real split  forms of  these groups,  which  are often
  denoted elsewhere by $E_{d+1|d+1}$ or $E_{d+1(d+1)}$.}.

One fruitful direction for investigating the nature of these constraints has been the study of terms in the low energy expansion of string theory amplitudes that generalise the amplitudes of classical supergravity.  For example, the four-supergraviton  amplitude for either of  the compactified type II string theories may be decomposed into the sum of the classical supergravity tree-level contribution,   an analytic part and a nonanalytic part,
 \be
 A_D(s,t,u)= A^{classical}(s,t,u)+ A_D^{analytic}(s,t,u) + A_D^{nonan}(s,t,u)\,,
 \label{ampsplit}
 \ee
 where the Mandelstam invariants  $s$, $t$, $u$ ($s+t+u=0)$ are quadratic in the momenta of the scattering particles\footnote{The Mandelstam invariants are $s=-(k_1+k_2)^2, t=-(k_1+k_4)^2, u=-(k_1+k_3)^2$, where $k_r$ ($r=1,2,3,4$) is the null momentum of particle $r$.}. The classical supergravity contribution that follows from the Einstein--Hilbert action can be written as
 \be
 A^{classical}(s,t,u) = \frac{3}{\sigma_3}\, \cR^4\,,
 \label{e:classical}
 \ee
 while the analytic part has a low energy expansion in powers of $s$, $t$, $u$, of the form
  \be
 A_D^{analytic}(s,t,u) = \sum_{p,q=0}^\infty\cE^{(D)}_{(p,q)}(\phi_{E_{d+1}/K})\, \sigma_2^p\, \sigma_3^q\, \cR^4 \,,
 \label{analytic}
 \ee
 where  $3\le D=10-d \le 10$, $\sigma_n=(s^n+t^n+u^n)\, (\ell_D^2/4)^n$, and $\ell_D$ is the $D$-dimensional Planck length.
  The term  ``supergraviton'' refers to the 256  massless physical states
 of  the maximal  supergravity multiplet,  which  have superhelicities
 that  enter in  the generalised  curvature tensor, $\cR$.  The four powers of this tensor in the kinematic factor
 $\cR^4$  are contracted by  a rank-sixteen  tensor, which  is defined
 in~\cite{Green:1987sp}.  Since $s$,  $t$  and $u$  are quadratic  in
 momenta,  a   term  of  the  form   $\sigma_2^p\sigma_3^q  \,  \cR^4$
 contributes a term  in an effective action of  the form $\partial^{4p
   +6q}\, \cR^4$ (where the derivatives are contracted into each other
 in  a standard manner).    So the infinite series of higher momentum terms translates into a series of higher derivative local interactions in an effective action that generalises the Einstein--Hilbert action.
The nonanalytic contribution, $A^{nonan}_D$, contains threshold singularities in $s,t,u$  which depend on the dimension, $D$.  Although there is generally no unambiguous way of disentangling these from the analytic part, this issue does not affect the terms of low order that are the concern of this paper. Nevertheless, even in the simplest cases the known threshold structure provides strong constraints on the behaviour of the coefficients in \eqref{analytic} near the cusp at which the $d$-torus decompactifies to the ($d-1$)-torus  \cite{Green:2010wi}, which will be of importance later.

 \begin{figure}[ht]
 \centering\includegraphics[width=8cm]{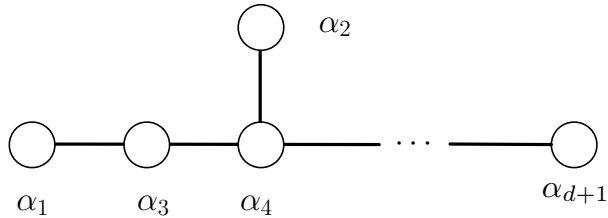}
 \caption{\label{fig:dynkin}  The  Dynkin  diagrams of  the  U-duality
   groups real split form of rank $d+1$ Lie group $E_{d+1}$ ($0\leq d\leq 7$)}
 \end{figure}

 The  duality symmetry of string theory implies that the $D$-dimensional
 amplitude should be invariant  under the action of the duality group,  $E_{d+1}(\mathbb Z)$.  As a consequence,
the coefficient functions, $\cE^{(D)}_{(p,q)}(\phi_{E_{d+1}/K})$, in~\eqref{analytic} must be automorphic functions  of  the symmetric  space moduli,  $\phi_{E_{d+1}/K}$, that parameterise the coset space $E_{d+1}/K$ appropriate to compactification  on $\calT^d$, where $K$ is
the  maximal compact  subgroup of  the duality group $E_{d+1}$. The  list of  the various
duality groups is given  in table~\ref{tab:Udual}.
 These moduli-dependent coefficient functions  contain a wealth of information relating perturbative and non-perturbative string theory effects.

   Although the structure of generic coefficients appears to be highly intractable,   the first three terms, for which $2p+3q\le 3$, are expected to display simplifying features as a consequence of maximal supersymmetry.  These  three interactions preserve a fraction of the complete 32-component supersymmetry, and should therefore be described as ``F-terms'', or fractional BPS interactions.  To be explicit,  the interaction $\calE^{(D)}_{(0,0)}\, \cR^4$ is 1/2-BPS,  $\calE^{(D)}_{(1,0)}\, \partial^4\,\cR^4$ is 1/4-BPS and $\calE^{(D)}_{(0,1)}\, \partial^6\, \cR^4$ is 1/8-BPS.    A BPS condition on an interaction generally implies that it is protected from receiving perturbative contributions beyond a certain order.  In other words,  such functions should  have a finite number of power-behaved terms when expanded around their cusps. They should also have a calculable spectrum of instanton, or non-zero mode,  contributions.   However, it is notoriously difficult to determine the extent of the constraints imposed on systems with maximal supersymmetry due to the absence of a covariant off-shell formulation.
 The  next term  in the  expansion,  $\calE^{(D)}_{(2,0)}\, \partial^8
 \cR^4$,  is expected to  be non-BPS~\cite{Green:2010},  and therefore
 not  protected by  supersymmetry, in  which case  its  coefficient is
 likely to have an infinite number of perturbative terms (power-behaved components in its expansion around any cusp).

Although the coefficients, $\calE^{(D)}_{(p,q)}$, have not been determined in generality,
a significant amount of information has accumulated for the first three terms for the cases with $D\ge 6$ (duality groups $E_{d+1}$ with $d \le 4$)~\cite{Green:1997tv, Green:1997as, Russo:1997mk,Green:2005ba,  Kiritsis:1997em,Basu:2007ru, Basu:2007ck},  and there are various conjectures concerning  the coefficient of the $\cR^4$ interaction,  $\calE^{(D)}_{(0,0)}$, for higher-rank  groups~\cite{Obers:1999um,Green:2010wi,Bao:2007er,Lambert,Pioline:2010kb}.   In addition, there are partial results for $2p+3q\le 6$ in $D=9$ dimensions with duality group $SL(2)$~\cite{Green:2008bf}.

The structure of the coefficients is highly constrained by a combination of string theory and M-theory input, which provides asymptotic information at various cusps in the space of moduli, together with
an analysis of the constraints imposed by supersymmetry~\cite{Green:1998by}.  Extending this to the exceptional groups, relevant to the theory in $D=3,4,5$ dimensions,  requires more sophisticated techniques, which we will develop in this paper.
  In particular, the coefficients $\calE_{(0,0)}^{(D)}$ and $\calE_{(1,0)}^{(D)}$
satisfy the Laplace eigenvalue equations~\cite{Green:2010wi},
\begin{eqnarray}
  \left( \Delta^{(D)}                  -{3(11-D)  (D-8)\over
    D-2}\right)\,\cE^{(D)}_{(0,0)}&=&  6\pi\,\delta_{D-8,0} \,,
\label{laplaceeigenone}\\
\left( \Delta^{(D)}  -{5(12-D) (D-7)\over D-2}\right)\,\cE^{(D)}_{(1,0)}&=&40\zeta(2)\, \delta_{D-7,0} \,,
\label{laplaceeigentwo}
\end{eqnarray}
where $\Delta^{(D)}$  is the Laplace  operator on the  symmetric space
$E_{d+1}/K$.  The Kronecker $\delta$ contributions on the
right-hand-side of these equations arise from anomalous behaviour, indicating the presence of polar terms for specific values of $D$ for which the eigenvalues in~\eqref{laplaceeigenone} and~\eqref{laplaceeigentwo} vanish.
The coefficient  $\calE^{(D)}_{(0,1)}$ satisfies the inhomogeneous Laplace eigenvalue equation
\begin{equation}
\left( \Delta^{(D)} -{6(14-D) (D-6)\over D-2} \right)\,\cE^{(D)}_{(0,1)}=-\left(\cE_{(0,0)}^{(D)}\right)^2 + 120\zeta(3)\,  \delta_{D-6,0}\,,
\label{laplaceeigenthree}
\end{equation}
which involves a source term on the right-hand side that is quadratic in the $\cR^4$ coefficient $\calE^{(D)}_{(0,0)}$.
The origin of~\eqref{laplaceeigenone}-\eqref{laplaceeigenthree}
 and, in particular, the values of the eigenvalues  on the left-hand sides of these equations
 was discussed in appendix~H of~\cite{Green:2010wi}.

Automorphic functions of moderate growth (which we assume ours are because of physical constraints) are nearly determined by imposing boundary
conditions  that specify  the  behaviour of  their  constant terms  in
various maximal parabolic subgroups that arise at boundaries of moduli
space; the only possibility ambiguity is an additive cusp form.  The constant  terms are  zero  Fourier modes  with respect  to
integration  over  the  unipotent   radical,  $N$,  in  the  Langlands
decomposition $P=MN$ of a parabolic  subgroup, $P$, where $M$ is its
Levi factor.
The  expressions  for  the   constant  terms  corresponding  to  three
particular   maximal  parabolic   subgroups  were   derived  explicitly
in~\cite{Green:2010wi} for the $E_{d+1}$  with $0\leq d\leq 4$ and for
$5\leq d\leq 7$ will be derived in this paper.

The Levi component $M$
 has the form $GL(1)\times G_d$, where $G_d$
is a rank  $d$ subgroup that corresponds to  deleting nodes $d+1$, $1$
or $2$ from the Dynkin in figure~\ref{fig:dynkin}, and is
given in table~\ref{tab:Parabolics}.  Such constant
terms contain a finite sum  of  components, of the
   form  $\sum_i r^{p_i}  \calF_{p_i}^{G_d}$, where $p_i$ are rational numbers,  $r$ is
the parameter for the $GL(1)$ factor  defined in~\ref{sec:description}, and the coefficients
$\calF_p^{G_d}$ are  automorphic functions for the subgroup $G_d(\IZ)$.
However, for the three subgroups appropriate to the string theory and
supergravity    calculations,  the coefficients are expected to be maximal parabolic Eisenstein series.
The behaviour at these boundaries was    discussed    in    detail    in
\cite{Green:2010wi,Green:2010}, and is summarised as follows.

\begin{itemize}
\item[(i)] The  subgroup obtained by removing  the root $\alpha_{d+1}$
  associated with the last node $d+1$ of the Dynkin diagram.  This is the ``decompactification limit'' in~\cite{Green:2010wi}, in which the radius of one compact dimension, $r_d/\ell_{D+1}=r^2$, becomes infinite, where $\ell_D$ is the $D$-dimensional Planck length.
 In this  case the parabolic subgroup  has a Levi factor,  $M$, of the
 form  $GL(1)\times  E_d$  and  the  constant  term  for  any  of  the
 coefficient functions  is a sum  of a finite  number of terms  of the
 form $\sum_i r^{p_i} \calF_{p_i}^{E_d}$ (suppressing some factors of $\log r$), where $\calF_{p_i}^{E_{d}}$ is an automorphic function for the group $E_d$.  This leads to a chain of relations from which it is possible to deduce all of the Eisenstein series from the $E_8$ case,\vspace{-1cm}
 $$
\qquad \quad E_8 \supset E_7 \supset E_6 \supset SO(5,5) \supset SL(5) \supset   SL(3)\times SL(2) \supset    SL(2)\vspace{-1cm}
 $$
 \item[(ii)]  The subgroup  obtained by  removing the  root $\alpha_1$
   associated with the node 1. This is the ``string perturbation theory
   limit'' in~\cite{Green:2010wi}, in which the  amplitude is expanded
   for  small  string  coupling,   $y_D=r^{-4}$.   In  this  case  the
   parabolic subgroup has a Levi  factor of the form $GL(1)\times SO(d,d)$
   and the constant term  is a sum of a finite number  of terms of the
   form  $\sum_p r^p  \calF_p^{SO(d,d)}$.   These correspond  to terms  in
   perturbative string theory, which  have values that can be obtained
   by  explicit integration  over string  world-sheets embedded   in
   $\calM^{10-d}\times     \calT^{d+1}$,     where    $\calM^D$     is
   $D$-dimensional Minkowski  space.

     \item[(iii)]   The  subgroup  obtained   by  removing   the  root
       $\alpha_2$ associated to the node 2.  This is the limit in which the volume of the M-theory torus, $\cV_{d+1}/\ell_{11}^{d+1}= r^{(2+2d)/ 3}$, becomes large.
     In this case the parabolic subgroup has a Levi factor of the form $GL(1)\times SL(d+1)$ and the constant term is a sum of a finite number of  terms of the form $\sum_p r^p \calF_p^{SL(d+1)}$.
      In   this  limit the semi-classical  approximation to eleven-dimensional supergravity is a good approximation and the values of the constant terms can be determined by evaluating one and two-loop Feynman diagrams embedded in $\calM^{10-d}\times \calT^{d+1}$.
   \end{itemize}

 Detailed knowledge of these boundary conditions is nearly sufficient to
   determine     the     solutions    to
   equations~\eqref{laplaceeigenone}-\eqref{laplaceeigenthree}.   With  these  boundary conditions, we show that the solutions of~\eqref{laplaceeigenone} and~\eqref{laplaceeigentwo} are sums of Eisenstein series defined with respect  to specific parabolic subgroups of the group $E_{d+1}$ -- up to the possible additive ambiguity of cusp forms.\footnote{Cusp forms often arise as ``error terms'' in arithmetic expansions, dating back to the classical function $r_4(n)$, the number of ways an integer $n$ can be written as the sum of 4 squares. The generating function $\sum_{n\ge 0}r_4(n)e^{2\pi i nz}=(\sum_{n\in\Z}e^{2\pi i n^2z})^4$, and hence is a modular form of weight 2 for the congruence subgroup $\G_0(4)$. There are no cusp forms of that weight for this group, and so the generating function is exactly an Eisenstein series -- resulting in striking identities, such as $r_4(n)=8(n+1)$ for $n$ prime. However, as one looks at sums of more squares and the weight increases, cusp forms inevitably creep in and complicate the formulas. Our situation is similar:~the existence of cusp forms would add a surprising touch of complexity to the Fourier coefficients and asymptotics of the solutions to~\eqref{laplaceeigenone}-\eqref{laplaceeigenthree}.}  This was  demonstrated in detail in~\cite{Green:2010wi} for $D\ge 6$, i.e,  $d\le  4$, and is generalized here to $d\le 7$.  Such cusp forms seem unlikely on purely mathematic grounds, because they have small Laplace eigenvalues. For example, the eigenvalues in~\eqref{laplaceeigenone}-\eqref{laplaceeigenthree} have the wrong sign to be part of the cuspidal spectrum unless $D$ is small enough. Even when the sign is correct, the papers ~\cite{Miller1996,Miller2001} give  lower bounds on the cuspidal Laplace spectrum on $SL(n,\Z)\backslash SL(n,\RR)/SO(n,\RR)$, for any $n$, which rule out such eigenvalues on this quotient.    It seems plausible Langlands functorial lifting from the $E_{d+1}$ groups to $SL(n,\RR)$ could (at least conjecturally) reduce our cases of interest here to the results of~\cite{Miller1996,Miller2001}. Such a link would however require serious technical sophistication, and is beyond the scope of this paper.

 In this paper we will extend this analysis to the remaining cases  $5\le d \le 7$, relating to $E_6$, $E_7$ and $E_8$.  This involves a detailed analysis of  constant terms of Eisenstein series for these groups, which will be the subject of section~\ref{sec:eisenseries}.  The general analysis leads to very large numbers of power-behaved components in the constant terms.  However, for the very special Eisenstein series of relevance to the string theory considerations there are immense simplifications and the relevant constant terms take the simple form expected according to items (i), (ii) and (iii).
  The application of these results into string theory language will be the subject of section~\ref{sec:detaildegen}.
  There, it will be seen that there is precise agreement between the values of the constant terms and the expectations based on string theory.

  The complete expressions for the constant terms of relevance are contained in a number of tables in the appendix.

  The solutions of~\eqref{laplaceeigenthree} are more general automorphic functions, $\calE^{(10-d)}_{(0,1)}$.  Their constant terms contain exponentially suppressed terms as well as terms that are powers of $r$ and were analysed for  $d\le 3$ in~\cite{Green:2005ba,Basu:2007ck,Green:2010wi}, and for the case $d=4$ in appendix~A of~\cite{Green:2010}.  The relevant constant terms for these coefficients in the $E_8$ case will be determined in section~\ref{sec:inhomogeneous}.
 As we will show, the power-behaved components of the constant term in the decompactification limit (i)  for the $E_8$ case contain  within them all three   of the $E_7$ coefficients, $\calE^{(4)}_{(0,0)}$, $\calE^{(4)}_{(0,1)}$ and $\calE^{(4)}_{(0,1)}$.

\section{Eisenstein series, parabolic subgroups and their constant terms}
\label{sec:eisenseries}

This section contains an introduction to Langlands Eisenstein series on higher rank groups~\cite{Langlands} and a description of some of his main results, followed by a computation of their constant terms in maximal parabolics.  The discussion is geared towards the relevant setting of this paper, though we also make an effort to explain more general phenomena that may later be useful for string theorists.  In particular we mainly curtail the discussion to two particular types of Eisenstein series:~{\em minimal parabolic Eisenstein series}, and {\em maximal parabolic Eisenstein series induced from the constant function} (which we shall see are specializations of the former).

We follow Langlands' Euler Products manuscript~\cite{Langlands-eulerproducts} in restricting to split Chevalley groups, as these are the only ones which arise in our investigations.  In fact, we only need to study simply laced ones, i.e., either G equals $SL(n)$, $SO(n,n)$, or a split form of $E_6$, $E_7$, or $E_8$.
  Let $B$ denote a fixed minimal parabolic ``Borel'' subgroup of $G$.  We decompose $B=MN$, where $M$ is its Levi component and $N$ its unipotent radical.
The Cartan subgroup of $G$ shall be denoted by $A$.

\subsection{Eisenstein series in classical terminology}

Researchers in automorphic forms typically define Eisenstein series in terms of adele groups because of the computational benefits this framework affords.  However, this is not necessary  to state the definitions.  In the present work it is important to understand the connection between Eisenstein series and other lattice constructions common in string theory.  Hence we felt it appropriate to define the series in concrete terms, which we shall do  in this subsection before recasting the definitions adelically in the next one.

Let us now consider the real points $G(\RR)$ of $G$, and let $\Delta$ denote the roots of $G(\RR)$ relative to the Cartan $A(\RR)$.  For each root $\a\in\Delta$, let $X_\a$ denote the Chevalley basis vector in the Lie algebra $\frak g$ of $G(\RR)$ that represents it, and $n_\a(t)=e^{tX_\a}$ the one-parameter unipotent subgroup it generates.  Furthermore we may form the Cartan Lie element $H_\a=[X_\a,X_{-\a}]\in {\frak a}$, the Lie algebra of $A(\RR)$.  If $\Sigma^+ \subset \Delta$ denotes the positive simple roots, then  $\{H_\a \,|\, \a\in \Sigma^+\}$ spans $\frak a$.  Thus we may identify elements of the connected component $A(\RR)^0$ of $A(\RR)$ with the exponentials $e^{\sum_{\a\in\Sigma^+}c_\a H_\a}$,  each $c_\a$ ranging over $\RR$.  The Iwasawa decomposition of $G(\RR)$ states that its elements $g$ each have unique decompositions $g=nak$, with $n\in N(\RR)$, $a\in A(\RR)^0$, and $k\in K$, a maximal compact subgroup of $G(\RR)$.  Thus there is a well-defined map $H:G(\RR)\rightarrow {\frak a}$ such that $g\in Ne^{H(g)}K$.

The roots $\a\in\Delta$ are by definition linear functionals on $\frak a$, and every linear functional $\l \in \frak a^*\otimes \C$ is  a linear combination of elements of $\Sigma^+$ with complex coefficients.  In what follows it is helpful to normalize definitions using the linear functional $\rho$,   defined to be  half the sum of all positive roots.  We denote the pairing between $\frak a^*\otimes \C$ and $\frak a$ by $\langle \cdot , \cdot \rangle$.  When $H_\a$ and $\a$ are tacitly identified, this corresponds to the usual inner product for the root system.
The Weyl group $\Omega$ acts both on $\frak a$ and  dually on  $\frak a^*\otimes \C$ in a way which preserves $\langle \cdot,\cdot \rangle$, and can be explicitly identified through any realization of the root system.

The function $H(g)$ is visibly unchanged if $g$ is multiplied on the left by an element of $N(\RR)$, and in particular any element of $N(\Z)=N(\RR)\cap G(\Z)$, where $G(\Z)$ is defined as in~\cite{Mizoguchi:1999fu}, or equivalently as  the stabilizer in $G(\RR)$ of the lattice spanned by the Chevalley basis~\cite{Chevalley}.  It is likewise invariant under $A(\Z)=A(\RR)\cap G(\Z)$ (because this finite group is contained in $K$), and hence under $B(\Z)=B(\RR)\cap G(\Z)=N(\Z)A(\Z)$ as well.  Eisenstein series are formed by averaging such objects over cosets of a group modulo a subgroup it is invariant under:

\begin{definition}
 The {\bf minimal parabolic Eisenstein series}  for $G$  is the coset sum
\begin{equation}\label{minparabZ}
    E^G(\l,g) \ \ := \ \ \sum_{\g \in B(\Z)\backslash G(\Z)} e^{\langle \l+\rho, H(\g g)\rangle }\,.
\end{equation}
This sum is absolutely convergent when the real part of $\l \in \frak a^*\otimes \C$ has sufficiently large inner products with all $\a \in \Sigma^+$.
\end{definition}

\noindent It is  a famous result of Langlands  that it meromorphically
continues to all of $\frak  a^*\otimes \C$, to an automorphic function
on  $G(\Z)\backslash  G(\RR)$.   When  $G=SL(2,\RR)$  this  definition
recovers           $\f{1}{2\,\zeta(2s)}\sum_{(m,n)\,\in\,\Z^2           -
  (0,0)}(\f{y}{|m\tau+n|^2})^s$, the usual non-holomorphic Eisenstein series for $SL(2,\Z)$.  One can of course trivially modify the definition to apply to subgroups $\G\subset G(\Z)$, though this appears to be unnecessary for our investigations.

The power functions $e^{\langle \l+\rho, H(\g g)\rangle }$ and hence $E^G(\l,g)$ itself  are always eigenfunctions of the Laplacian:
\begin{equation}\label{e:LaplaGeneral}
  \Delta^{G/K}\,           E^G(\lambda,g) \ \ = \ \            2\,(\langle
  \lambda,\lambda\rangle-\langle
  \rho,\rho\rangle)\, E^G(\lambda,g)\,.
\end{equation}
This  formula  for  the  eigenvalue  is crucial  for  identifying  the
solutions to~\eqref{laplaceeigenone}-\eqref{laplaceeigenthree}, and is
completely   analogous  to  the   $SL(2)$  fact   that  $y^s$   is  an
eigenfunction  of   the  hyperbolic   Laplacian.   It  is   proven  by
identifying $\Delta^{G/K}$  as a multiple  of the Casimir on  $G$, and
using explicit formulas for the latter (see~\cite[p.~303]{Knapp1988}).
Actually  the   power  functions  and  hence   Eisenstein  series  are
eigenfunctions of  not merely  $\Delta^{G/K}$, but furthermore  of the
full ring of invariant differential operators.  This is a crucial fact
in  Langlands'  meromorphic  continuation.   So far  string  theoretic
arguments  have mainly produced  information about  $\Delta^{G/K}$ and
not   these  other   operators,   so~(\ref{e:LaplaGeneral})  naturally
determines only  $\langle \lambda,\lambda \rangle$.   The structure of
the constant terms and the integrality constraint discussed in~\secref{sec:3.1} is then used to pin down $\l$ exactly there.

We have used the term ``minimal parabolic'' series for these because of the r{\^o}le the Borel subgroup $B$ plays in their definition.  In general, any subgroup $P$ that contains $B$ is called a  {\em standard parabolic subgroup}; $P$ is called a {\em maximal parabolic} if $G$ itself is the only subgroup that properly contains it.  All parabolic subgroups have the unique decomposition $P=M_PN_P$, where $M_P$ is its Levi component and $N_P$ its unipotent radical. The standard parabolics of $G$ are in one-to-one correspondence with subsets $S \subset \Sigma^+$ as follows:~$M_P$ includes all $n_{\pm \a}$ for $\a\notin S$, while $N_P$ contains all $n_\a$ for $\a\in S$.  In particular, each maximal parabolic subgroup is associated to a single, simple root $\b$, and we shall sometimes use the notation $P=P_\beta$ to emphasize this dependence.

Another family of important Eisenstein series that arise in our string theory calculations are {\em maximal parabolic} Eisenstein series.  Let us first explain the simplest versions, which are induced from constant functions -- they are in fact generalizations of the classical Epstein Zeta functions.
  These series are formed in a similar way to (\ref{minparabZ}), but with special parameters $\l$ such that $\langle \l + \rho , H(g)\rangle$ is unchanged if $g$ is multiplied on the left by an element of $P(\Z)$, where $P\supset B$ is a designated standard maximal parabolic subgroup of $G$.
  This is equivalent to requiring  $\l+\rho$ be orthogonal to any simple root $\alpha$ other than the one $\beta$ which defines the maximal parabolic $P=P_\beta$, and restricts $\l$ to lie on a line in $\frak a^*\otimes \C$.  In terms of the dual basis  $\{\omega_\a|\a\in\Sigma^+\}$ defined by the condition that $\langle \omega_\a,\beta \rangle = \d_{\a=\beta}$, these $\l$ can be parametrized in terms of single complex variable $s$ as
\begin{equation}\label{lambdaparam}
    \lambda \ \ = \ \ 2\,s\,\omega_\beta \ - \ \rho\,.
\end{equation}
The following definition uses this special choice of $\l$, but restricts the range of summation owing to the extra invariance of the summand under $P(\Z)$:
\begin{definition}\label{maxparabdef}
For  $P=P_\beta$, the {\bf maximal parabolic Eisenstein series induced from the constant function} is
\begin{equation}\label{maxparabZ}
   E^G_{\beta;s} \ \ := \ \ \sum_{\g \in P(\Z)\backslash G(\Z)} e^{\,2\,s\,\langle  \omega_\beta, H(\g g)\rangle }\,.
\end{equation}
\end{definition}

\noindent Our normalization of $s$ is chosen so that it agrees with the usual non-holomorphic Eisenstein series for $SL(2,\Z)$.  These series of course also have meromorphic continuations to $s\in \C$, and specialize to be identically equal to 1 at the special point $s=0$ because of the following fact (whose proof we shall describe later):
\begin{thm}\label{valueatminusrho}
    The minimal parabolic Eisenstein series $E^{G}(-\rho,\cdot)$ equals the constant function 1.
\end{thm}

To connect these two definitions, it is worthwhile to consider yet another type of maximal parabolic Eisenstein series that features  another ingredient:~an additional factor $\phi(\g g)$ in the summand (\ref{maxparabZ}) which is an automorphic function on the Levi component of $P$, extended to $G$.  Such a sum is still well-defined and has similar convergence properties. Definition~\ref{maxparabdef} amounts to setting this function equal to 1.  Interestingly, the inclusion of this function allows us to view the maximal parabolic Eisenstein series (\ref{maxparabZ}) as a special case of (\ref{minparabZ}).  Indeed, recall the decomposition $P=M_PN_P$ from above, where $M_P$ is its Levi component and $N_P$ its unipotent radical.  The intersection $B_P:=M_P\cap B$ is itself the Borel subgroup of the reductive group $M_P$, and the coset representatives $\g\in B(\Z)\backslash G(\Z)$ can be uniquely decomposed into as products $\g=\g_1\g_2$ with $\g_1\in B_P(\Z)\backslash M_P(\Z)$, $\g_2\in P(\Z)\backslash G(\Z)$.  Hence we may write (\ref{minparabZ}) as the double sum
\begin{equation}\label{minparabZdoublesum1}
      E^G(\l,g) \ \ := \ \ \sum_{\g_1\in B_P(\Z)\backslash M_P(\Z)} \sum_{\g_2\in P(\Z)\backslash G(\Z)} e^{\langle \l+\rho, H(\g_1 \g_2 g)\rangle }\,.
\end{equation}

Any element $\l \in {\frak a}^*\otimes \C$ can be uniquely decomposed as $\l=\l^P+\l_P$, where $\l^P$ is a complex linear combination of simple roots not equal to $\beta$ (the root defining $P=P_\beta$), and $\l_P$ is orthogonal to all such simple roots.  This decomposition in particular applies to $\rho$, expressing it as the sum of $\rho^P$ (which is itself half the sum of the positive roots of $M_P$), and $\rho_P$ (which is a scalar multiple of $\omega_\beta$).    Using this decomposition, we write the exponent as
\begin{equation}\label{exponent}
\aligned
    \langle \l+\rho, H(\g_1 \g_2 g)\rangle \ \ & = \ \  \langle \l_P+\rho_P, H(\g_1 \g_2 g)\rangle  \ + \ \langle \l^P+\rho^P, H(\g_1 \g_2 g)\rangle
     \\
     & = \ \ \langle \l_P+\rho_P, H( \g_2 g)\rangle  \ + \ \langle \l^P+\rho^P, H(\g_1 \g_2 g)\rangle\,;
\endaligned
\end{equation}
in the last step we have used the fact that $H(mg)$ and $H(g)$ have the same inner product with $\l_P+\rho_P$, for any $m$ generated by the $n_{\pm \a}$ for $\a \in \Sigma^+_{\neq \beta}$, in particular $M_P(\Z)$.
 Hence
 (\ref{minparabZdoublesum1})  can be expressed in the range of absolute convergence as
\begin{equation}\label{minparabZdoublesum2}
      E^G(\l,g) \ \ := \ \  \sum_{\g_2\in P(\Z)\backslash G(\Z)} e^{\langle \l_P+\rho_P, H( \g_2 g)\rangle }\,\phi(\g_2 g)\,,
\end{equation}
where $\phi(g):=\sum_{\g_1\in B_P(\Z)\backslash M_P(\Z)} e^{\langle \l^P+\rho^P , H(\g_1 g) \rangle }$ is now a minimal parabolic Eisenstein series for the smaller reductive group $M_P$.  Thus minimal parabolic Eisenstein series are themselves special cases of maximal parabolic Eisenstein series -- but induced from the function $\phi$ rather than the constant function.

We can now see that~(\ref{maxparabZ}) is a specialization that coincides with (\ref{minparabZ}) when $\l$ has the form (\ref{lambdaparam}).  In this case $\l^P=-\rho^P$, and the Eisenstein series on $M_P$ in the previous paragraph specializes to be constant because of Theorem~\ref{valueatminusrho}.  Hence under the special assumption (\ref{lambdaparam}), the inducing function $\phi$ is constant and the two notions coincide.

\subsection{Eisenstein series in adelic terminology}

We have just given definitions of the Eisenstein series involved in this paper, in concrete classical terms. Adele groups are often used in automorphic forms as a notational simplification that hints to effective ways to group terms together in calculations.  In the context of Eisenstein series, they are used to  reparametrize the sums over $P(\Z)\backslash G(\Z)$ (whose cosets can be intricate to describe).  This application --  a brilliant insight of Piatetski-Shapiro that was furthered by Langlands -- has been crucial in readily obtaining exact formulas (by comparison, Poisson summation is much harder to execute directly).  Since this is crucial to our calculations, we have elected to give a description here.

Let us return to (\ref{minparabZ}) and its sum over cosets $B(\Z)\backslash G(\Z)$.  As we mentioned above, cosets for this quotient can be difficult to directly describe, especially as the group $G$ gets more complicated.  This is because $\Z$ is a ring, not a field like $\RR$ (where the corresponding quotient is just the maximal compact subgroup $K$).  Strikingly, the field $\Q$ gives the same coset space as $\Z$: the inclusion map from $G(\Z)$ into $G(\Q)$ induces the bijection of cosets
\begin{equation}\label{BZBQ}
    B(\Z)\backslash G(\Z) \ \ \simeq \ \ B(\Q)\backslash G(\Q)\,.
\end{equation}
This is because our assumptions on $G$ imply that  $G(\Q)=B(\Q)G(\Z)$  (see~\cite[\S 2]{Langlands-eulerproducts}).   Similarly, $P(\Z)\backslash G(\Z)\simeq P(\Q)\backslash G(\Q)$, and therefore our Eisenstein series can in principal be written as sums over these rational quotients.  To do this properly one needs to redefine $H$ in such as away that it is invariant under $B(\Q)$ on the left.  Note that the existing definition does not qualify, because $A(\Q)$ is dense in $A(\RR)$ and hence $H$ cannot be trivial on it.

The remedy is to instead consider the each of the groups $G(\Q_p)$,  where $p$ denotes either a prime number or $\infty$ (in the latter situation, we follow the convention that $\Q_\infty=\RR$).  Just as the adeles $\A$ are the restricted  product of all $\Q_p$ with respect to the $\Z_p$ (i.e., all but a finite number of components of each element lie in $\Z_p$), the adele group $G(\A)$ is the restricted  product of each $G(\Q_p)$ with respect to the $G(\Z_p)$ = the stabilizer of the Chevalley lattice, tensored with $\Z_p$.

  A variant of the Iwasawa decomposition persists in the $p$-adic case as well: $G(\Q_p)=N(\Q_p)A(\Q_p)G(\Z_p)$, though this is no longer unique since, for example, $A(\Q_p)\cap G(\Z_p)$ is nontrivial.  Thus, globally, we have that $G(\A)=N(\A)A(\A)K_\A$, where $K_\A$ is the product of the real group $K$ with all $G(\Z_p)$'s. Since $G$ is assumed to be split, $A(\A)$ is a product of $r=rank(G)$ copies of the torus $GL(1,\A)=\A^*$, the ideles of $\A$.  Strong approximation for the ideles equates the quotient $\Q^*\backslash \A^*$ with $(\{\pm 1\} \backslash \RR^* )\times \widehat{\Z}^*$, where $\Q$ is regarded as {\em diagonally}\footnote{i.e., into each factor $\Q_p$ simultaneously.} embedded into $\A$, and $\widehat{\Z}^*$ is the product of all $\Z_p^*$'s.  Since the first factor is isomorphic to $\RR$ via the logarithm map, this identification extends $H$ from a function on $A(\RR)\simeq (\RR^*)^r$ to   an $A(\Q)$-invariant, $A(\widehat{\Z}^*)$-invariant function from $A(\A)$ to $\frak a$.  Furthermore, it extends to $G(\A)$ through the global Iwasawa decomposition to a left $B(\Q)$-invariant function, with $B(\Q)$ likewise thought of as being diagonally embedded into $G(\A)$.   We can now define the adelic Eisenstein series as
\begin{equation}\label{minparab}
   E^G(\l,g) \ \ = \ \ \sum_{\g \in B(\Q)\backslash G(\Q)} e^{\langle \l+\rho, H(\g g)\rangle }\,,
\end{equation}  which we stress agrees with (\ref{minparabZ}) when the argument $g\in G(\RR)$, and defines an extension to $g\in G(\A)$ that is left-invariant under the diagonally embedded $G(\Q)$.  Similarly
\begin{equation}\label{maxparab}
    E^G_{\beta;s} \ \ := \ \ \sum_{\g \in P(\Q)\backslash G(\Q)} e^{\,2\,s\,\langle  \omega_\beta, H(\g g)\rangle }\,,
\end{equation}
which is again a specialization of (\ref{minparab}) under the assumption (\ref{lambdaparam}).

\subsection{Constant term formulas}
\label{sec:constantterm}

In general the Fourier expansions of Eisenstein series are intricate to state, and are not presently known in full detail.  However, a simple natural part of them (defined as follows) have very explicit formulas due to Langlands, and are crucial to the analytic properties of these series.
\begin{definition}
The {\em constant term} of an automorphic form $\phi$ in a standard parabolic subgroup $P\supset B$ is given by the
integral
\begin{equation}\label{constterm}
    \int_{N_P(\Z)\backslash N_P(\RR)} \phi(ng)\,dn\,,
\end{equation}
where $N_P$ is the unipotent radical of $P$, and $dn$ is Haar measure on $N_P$ normalized to give the quotient $N_P(\Z)\backslash N_P(\RR)$ volume 1.  (Since $N_P$ is unimodular, $dn$ is simultaneously both a left and right Haar measure.)
\end{definition}

We conclude this section by stating the constant term formula for the minimal parabolic Eisenstein series $E^G(\l,g)$ in maximal parabolics.  When (\ref{lambdaparam}) holds, the specialization of the formula below gives the constant terms of the maximal parabolic series in Definition~\ref{maxparabdef} -- this will be very useful for our applications.
The formula involves the functions
\begin{equation}\label{Msl}
    M(w,\lambda) \ \ = \ \ \prod_{\srel{\a\,\in\,\Delta^+}{w\a\,\in\,\Delta^-}}c(\langle \l,\a\rangle) \ , \ \ c(s) \ = \ c(-s)^{-1} \ = \ \f{\xi(s)}{\xi(s+1)}\,,
\end{equation}
which arise from intertwining operators.    Here $w$ is an element of the Weyl group $\Omega$ of $G$, $\lambda \in {\frak a}^*\otimes \C$, and $\xi(s)$ is the completed Riemann $\zeta$-function $\pi^{-s/2}\Gamma(\f{s}{2})\zeta(s)$.  They satisfy the cocycle identity \begin{equation}\label{Mslcocycle}
    M(w_1 w_2 ,\l) \ \ = \ \ M(w_1,w_2\l)\,M(w_2,\l)\,.
\end{equation}
Let $P=P_\gamma$ be any maximal parabolic, and
let $\Omega_P$ denote the Weyl group of $M_P$, thought of as a subgroup of $\Omega$.  As above, we  decompose  any $\l \in {\frak a}^*\otimes \C$ as the sum $\l_P+\l^P$, where $\l_P$ is perpendicular to all simple roots aside from $\gamma$,  and $\l^P$ is a multiple of $\omega_\gamma$.

\begin{thm}\label{consttermthm}(Langlands' Constant term formula -- see~\cite[Proposition II.1.7.ii, p.92]{Moeglin})
\begin{equation}\label{consttermformula}
   \int_{N_P(\Z)\backslash N_P(\RR)} E^G(\l,ng)\,dn \ \ = \ \
    \sum_{w \,\in\,\Omega_P\backslash \Omega}  e^{\langle (w\l)_P+\rho_P,H(g)\rangle}M(w,\l)\, E^{M_P}((w \l)^P,g)\,.
\end{equation}
\end{thm}
Of course the last Eisenstein series $E^{M_P}$ must be thought of as a product of Eisenstein series on the different reductive factors of $M_P$ when the latter is not simple.  Moeglin-Waldspurger actually prove a slightly different statement, where the $M(w,\l)$ appear directly as intertwining operators which satisfy a composition law identically compatible to~(\ref{Mslcocycle}).  Using embedded $SL(2)$'s inside $G$ it is easy to see that these operators act there as the scalar (\ref{Msl}) when $w$ is a simple Weyl reflection, and therefore on the full  Weyl group as well.  Moeglin-Waldspurger also deal with adelic Eisenstein series, which here can be equated to their classical variants via the correspondence described in the previous subsection.  Also, the adelic constant term integration over $N_P(\Q)\backslash N_P(\A)$ there drops to $N_P(\Z)\backslash N_P(\RR)$ for the adelization of classical Eisenstein series (this is because of strong approximation for $\A$).  Strictly speaking, we have also used the fact that the Weyl group of a Chevalley group sits inside $G(\Z)$.

Formula (\ref{consttermformula}) is consistent with two other important identities about the minimal parabolic Eisenstein series:~the functional equation
\begin{equation}\label{fe}
    E^G(\l,g) \ \ = \ \ M(w,\l) \,
    E^G(w\l,g) \ , \ \ w\,\in\,\Omega\,,
\end{equation}
and Langlands' constant terms formula in the minimal parabolic
\begin{equation}\label{fullconstantterm}
    \int_{N(\Z)\backslash N(\RR)}
   E^G(\l,ng)\,dn \ \ = \ \ \sum_{w\,\in\,\Omega}e^{\langle w\l+\rho,H(g)\rangle}\,M(w,\l)\,.
\end{equation}
For example, when $\l=-\rho$ the inner product of $\l$ and any simple root is $-1$, a point at which $c(s)$ vanishes.  Any nontrivial Weyl group element flips the sign of at least one simple root, so that $M(w,\l)$ vanishes for all but the identity element $w\in \Omega$.  This means the constant term (\ref{fullconstantterm}) is identically one, consistent with Theorem~\ref{valueatminusrho}.  Moreover, it is not hard to deduce Theorem~\ref{valueatminusrho} from this calculation.  Indeed, the general theory of Eisenstein series describes how  the constant terms control the growth of Eisenstein series:~since this one is bounded, so is the full series.  It  is furthermore a Laplace eigenfunction with eigenvalue 0 when $\l=-\rho$, and hence it is constant.

As  noted above, because  $E^G$ specializes  to the  maximal parabolic
Eisenstein          series          under         (\ref{lambdaparam}),
Theorem~\ref{consttermthm}  provides  the   constant  terms  of  those
objects  as  well.   It  is  possible to  prove  those  formulas  more
directly,  without  reference  to  the  minimal  parabolic  Eisenstein
series, for example as is  argued for some of the relevant cases earlier in~\cite{GRS}.   However, we felt it
was  important  to  calculate  everything via  the  minimal  parabolic
Eisenstein series  for two  different reasons.  The  first is  that we
will   rely   heavily   on   special   identities   relating   various
$E^G_{\beta;s}$   via   (\ref{fe})   that   are   more   apparent   as
specializations,  rather than  as identities  of sums  over completely
different coset  spaces.  The  second reason is  that the constant term calculation in the proof of 
\cite[Theorem 2.3]{GRS} does  not carry over to all Eisenstein  series, because they
assert  that the  intersection  of  the Levi  component  of a  maximal
parabolic with the conjugate of  another parabolic is itself a maximal
parabolic subgroup of this Levi.  However, this is false in general --
including  for a number  of subtle  examples we  faced in  the present
work, such as the case of $G=E_6$ when $\b=\a_5$   and the constant
term is taken in the maximal parabolic corresponding to $\a_1$ (in the numbering of Figure~\ref{fig:dynkin}). We would like to make clear that this in no way affects the validity of the results in~\cite{GRS}, because the assertion is valid in the cases they study -- rather it only affects extensions of their work to different situations.  What it means is that non-maximal parabolic Eisenstein series can arise in some of the constant terms we study.  Strikingly, these series tend to vanish at the special points of interest.

Finally, in addition to terms vanishing because a simple root is flipped, sometimes Eisenstein series vanish for a more subtle reason:~all their constant terms vanish.  In such a case, the Eisenstein series is by definition a cusp form, which much itself vanish because Eisenstein series are orthogonal to all cusp forms. It follows by induction that this is the case if and only if (\ref{fullconstantterm}) vanishes, which can happen even at points where the individual $M(w,\l)$ are singular because of cancellation between terms.  This can be very tedious to computationally check, even with the methods of the paragraph below.  This vanishing, however, is ultimately responsible for many of our simple formulas for constant terms at special points.

\subsection{Brief description of explicit computations used later in the paper}\label{sec:description}

Theorem~\ref{consttermthm} gives explicit formulas for all constant terms we are after, since it is possible to enumerate the Weyl group, describe its action on $\l$, and calculate the exact factors $M(w,\l)$ from (\ref{Msl}).  In practice, the Weyl group is so large that it is difficult to have an {\it a priori} explanation of what the calculation will give.  Langlands noticed  already in his example of the rank two group  $G_2$  in~\cite[Appendix 3]{Langlands}
that many terms vanish at special points.  Indeed, such vanishing  is absolutely crucial to the physical conclusions we draw from the case of $E_8$ and its subgroups.  Mathematicians have studied similar vanishing in different settings, and giving explanations for it (e.g.~\cite{Kudla-Rallis,GRS}).

Unfortunately, it was not initially obvious what configurations of parabolics $P_\gamma$ and maximal parabolic Eisenstein series $E^G_{\beta;s}$ to investigate, and it became a practical necessity to have a fast way to obtain explicit constant terms for large swaths of examples
in order to unravel structures which were important to us from a string-theoretic point of view.  Economical mathematical explanations for the phenomena at hand were not as important as speed, especially in light of the complexity of the objects involved.
To get around this issue, we first precomputed which Weyl group elements $w$  flip a simple root $\a\neq\beta$.  For such $\a$, $\langle 2s\omega_\beta-\rho,\a\rangle = -1$, which forces the product (\ref{Msl}) to vanish.  These terms, which provide the vast majority, can hence be discarded from the constant term calculations.  The remaining terms are stored for later calculations.  For example, out of the  696,729,600  $E_8$ Weyl group elements, only 240 are needed when $\beta$ is the 8-th root of $E_8$, i.e. the one which is added on from $E_7$.  To lessen storage requirements, we worked with a factorization of the $E_n$ Weyl groups in terms of the $E_{n-1}$ Weyl group, and a fixed set of coset representatives.  The resulting computer calculations were then very efficient, and allowed us to explore the properties of a large number of cases and identify significant patterns.  Ultimately the constant term calculations were automated, only rarely taking  more than 30 seconds a piece

The constant term is  a function on the Levi component $M_P$ of the maximal parabolic $P=P_\gamma$, which  is the product of a one-dimensional group with all Chevalley groups whose Dynkin diagrams compose the connected components of the Dynkin diagram of $G$, once the node for the simple root $\gamma$  is deleted (note that this is not in general the same root $\beta$ that defines the Eisenstein series).  This second factor is the group that the Eisenstein series $E^{M_P}$ on the  right hand side of (\ref{consttermformula}) is defined on. The exponential factor $e^{\langle (w\l)_P+\rho_P,H(g)\rangle}$ multiplying it depends on the one-dimensional piece.  To parameterize it uniformly, factor $g\in M_P(\RR)$ as $tg_0$, where $t$ is in this one-dimensional piece and $g_0$ is in the product of smaller Chevalley groups.  By construction, $t$ acts trivially on the roots spaces spanned by the $X_\a$ contained inside $M_P$, and is uniquely determined by its eigenvalue on $X_\g$, which we parametrize as $2\log r$.  Hence the exponential factor in (\ref{consttermformula}) is a power of $r$, and the constant term is a polynomial with exponents depending on $s$ whose coefficients are lower-rank Eisenstein series.

{\bf Example: $G=E_8$, $\b=\a_1$, $\g=\a_2$.}

\noindent
Here all but  2,160 out of the 696,729,600 Weyl group elements give zero contribution, and the polynomial just mentioned is equal to
$420 r^{84}+8 r^{230-20 s}+168 r^{162-12 s}+280 r^{134-8 s}+8 r^{140-8 s}+56 r^{114-4 s}+280 r^{-4 (-27+s)}+70 r^{-16 (-12+s)}+8 r^{20 s}+168 r^{12 (2+s)}+8 r^{8 (6+s)}+56 r^{4 (17+s)}+280 r^{62+4 s}+280 r^{42+8 s}+70 r^{8+16 s}$, provided the Eisenstein series are suppressed (otherwise the formula is even more unwieldy).  At the special point $s=\f 32$ all but the two smallest powers of $r$ vanish, giving
\begin{equation}\label{E821at32}
    r^{30} \,E^{A_7}_{\alpha_1;\threeh} \ + \ \f{\xi(2)}{\xi(3)}\,r^{32}
\end{equation}
as the constant term.

\section{String theory amplitudes and their degeneration limits}
\label{sec:detaildegen}
The results of the previous section allow us to extend the analysis of the structure of the automorphic functions  that arise in the expansion of the string theory amplitude~\cite{Green:2010wi}  to a more general setting that includes the exceptional duality groups.

It is useful to translate the terms in the low energy expansion of the analytic part of the scattering amplitude,~\eqref{analytic}  into local terms in an effective action, so that the first three terms beyond classical Einstein theory in $D$ dimensions are
\begin{equation}
  \label{e:R4effec}
  S_{R^4}=\ell_D^{8-D}\,\int       d^Dx      \,      \sqrt{-G^{(D)}}\,
  \cE^{(D)}_{(0,0)}\, \cR^4\, ,
\end{equation}
and
\begin{equation}
  \label{e:D4R4effec}
  S_{\p^4R^4}=\ell_D^{12-D}\,\int       d^Dx      \,      \sqrt{-G^{(D)}}\,
  \cE^{(D)}_{(1,0)}\, \partial^4 \cR^4\,,
\end{equation}
and
\begin{equation}
  \label{e:D6R4effec}
  S_{\p^6R^4}=\ell_D^{14-D}\,\int       d^Dx      \,      \sqrt{-G^{(D)}}\,
  \cE^{(D)}_{(0,1)}\, \partial^6 \cR^4\,.
\end{equation}

We  will   first consider the solutions for the coefficients $\calE^{(D)}_{(0,0)}$ and $\calE^{(D)}_{(1,0)}$, which satisfy the Laplace eigenvalue equations~\eqref{laplaceeigenone} and~\eqref{laplaceeigentwo}.
 The discussion of the  automorphic coefficient $\cE^{(D)}_{(0,1)}$,  which satisfies the more elaborate equation~\eqref{laplaceeigenthree},   will  be deferred to
 section~\ref{sec:inhomogeneous}.

\bigskip\noindent
{\bf Note on conventions}
  \medskip

The solutions will involve linear combinations of Eisenstein series of the kind described in the last section.  In
describing the string theory results it will prove convenient to use a normalisation for
maximal parabolic Eisentein series that includes a factor of $2\zeta(2s)$, so we will define
\begin{equation}\label{e:normalisation}
  \bE^{E_{d+1}}_{[0^u\,1\,0^{d-u}];s}:= 2\zeta(2s) \, E^{E_{d+1}}_{\beta;s}\,,
\end{equation}
where  $[0^u\,1\,0^{d-u}]$ is  the  Dynkin label  associated with  the
simple root
 $\beta$  in the  definition~\eqref{lambdaparam}.   Furthermore, since
 the conventional $SL(2)$ Eisenstein series has a trivial Dynkin label
 it  will  be  written  as  $\bE^{SL(2)}_s$. For  the  $SL(5)$  
 non-Epstein  series   the  normalisation  differs  by   a  factor  of
 $\zeta(2s-1)$ with respect to the one used in~\cite{Green:2010wi,Green:2010}.

  The parameter $r$, defined in section~\ref{sec:description}, associated with the $GL(1)$ factor in section~\ref{sec:eisenseries}
translates  into distinct  physical parameters  in each  of  the three
degeneration  limits that correspond  to deleting  nodes $d+1$,  $1$ and $2$, respectively,
of the Dynkin diagram  in fig. \ref{fig:dynkin}.  These are summarised as follows:
\bea
&&{\rm Limit~(i)}\qquad\ r^2 = {r_d/ \ell_{D+1}}\,,   \qquad  (r_d= {\rm radius\ of\ circle})\,,\nn\\
&&{\rm Limit~(ii)} \qquad \  r^{-4} = y_D= {\rm string\ term\ constant} \,,\nn\\
&&{\rm Limit~(iii)} \qquad \  r^{{2+2d\over 3}} ={ \calV_{d+1}/ \ell_{11}^{d+1}} \,, \ \ {\calV_{d+1}}  = {\rm volume\ of\ M\ theory\ torus}\,.
\nn
\eea
The  $D$-dimensional string  coupling constant  is defined  by  $y_D =
g_s^2\, \ell_s^d/V_d$,  where $g_s$  is the $D=10$  IIA or  IIB string
coupling and $V_d$ is the volume of $\calT^d$ in string units.
The Planck scales in different dimensions that enter in~\eqref{e:R4effec} and~\eqref{e:D4R4effec} are related to each other and the string scale, $\ell_s$, by
\bea
\ell_D^{D-2} =\ell_{D+1}^{D-1}{1\over r_d} = \ell_s^{D-2} y_D\ \ {\rm for}\ D\le 10\,,\qquad\ell_{11}= g_A^{\third}\, \ell_s\, ,
\label{plancks}
\eea
where $g_A$ is the IIA string coupling.

\subsection{Solutions for the coefficients $\calE^{(D)}_{(0,0)}$ and $\calE^{(D)}_{(1,0)}$.}
\label{sec:3.1}

We will show  that  the  automorphic   coefficients  in~\eqref{e:R4effec} and~\eqref{e:D4R4effec} are  given  by the  simple  expressions,
\be
\calE^{(D)}_{(0,0)} =2\zeta(3)\, E^G_{\alpha_1;{3\over2}}:=\bE^{E_{d+1}}_{[1\, 0^d];\threeh}\,,
\label{e:R4ans}
\ee
and
\be
\calE^{(D)}_{(1,0)} =\zeta(5)\,E^G_{\alpha_1;{5\over2}}:= {1\over 2}\, \bE^{E_{d+1}}_{[1\, 0^d];\fiveh}\,,
\label{e:D4R4ans}
\ee
for $3\le D \le 5$ (or $7 \ge d \ge 5$).
Substituting the expression~\eqref{lambdaparam} for  $\lambda$ in terms of $s$ into equation~\eqref{e:LaplaGeneral}, it follows that the solutions~\eqref{e:R4ans} and~\eqref{e:D4R4ans}
 satisfy the Laplace eigenvalue equations~\eqref{laplaceeigenone}     and~\eqref{laplaceeigentwo}  with
$s=3/2$ and $s=5/2$ respectively.
We will shortly  show that these functions also  satisfy the requisite
boundary conditions in the three limits of interest.  An important general comment about these boundary conditions is that in each of the three limits the   automorphic
coefficients have moderate power-like growth for $r\to \infty $.
 Another necessary condition for these being acceptable solutions is that
in the limits~(i)  and limit~(ii) they give rise  to integer powers of
the    radius   $r_d/\ell_{D+1}=r^2$    and   the    string   coupling
$y_D=r^{-4}$.  Strictly speaking our calculations  merely show that (\ref{e:R4ans}) and (\ref{e:D4R4ans}) solve all relevant equations, but do not directly rigorously show that they are unique solutions.  However, any two solutions differ only by a linear combination of cusp forms and other Eisenstein series.  The constant terms of these series can be computed explicitly as well.  Though we do not fully investigate this here, the possibility of other solutions seems unlikely because of known non-existence results for cusp forms on $SL(n,\Z)\backslash SL(n,\RR)/SO(n,\RR)$ \cite{Miller1996,Miller2001}, the conjectured properties of Langlands' functorial lifts, and the rationality of the cuspidal eigenvalues the above integrality constraint  dictates.

First we will comment on the form of these solutions.  From the general expression~\eqref{lambdaparam}  for the weight vector that defines a maximal parabolic Eisenstein series, the vector associated with a maximal parabolic Eisenstein series $E^G_{\beta;s}$ is $\lambda_{\beta}(s)= 2 s \,\omega_{\beta} -\rho$.
For the  series in
\eqref{e:R4ans} and~\eqref{e:D4R4ans} this has the form $\lambda_{\alpha_1}(s)= 2 s \,\omega_{\alpha_1} -\rho$, where $\omega_{\alpha_1}$ is the weight vector for the simple root $\alpha_1$ labelling the first node of the Dynkin diagram in figure~\ref{fig:dynkin}.
Therefore, the weight vector associated with the series $E^{E_{d+1}}_{\alpha_1;3/2}$ is $\lambda_{\alpha_1}(3/2)$, while for the series $E^{E_{d+1}}_{\alpha_{d+1};5/2}$ it is $\lambda_{\alpha_1}(5/2)$.  As we know from earlier examples, there are many equivalent ways of expressing the same series as those in~\eqref{e:R4ans} and~\eqref{e:D4R4ans}.
For  the  exceptional groups,  $E_{d+1}$  with  $d=5,6,7$, the  weight
vectors   $\lambda_{\alpha_1}(3/2)$,   $\lambda_{\alpha_{2}}(1)$   and
$\lambda_{\alpha_{d+1}}((d-2)/2)$  are  in the  same  orbit under  the
action     of     the     Weyl     group,     $\Omega$.     Similarly,
$\lambda_{\alpha_1}(5/2)$  and  $\lambda_{\alpha_{d+1}}((d+2)/2)$  are
also in the same Weyl orbit.   This means that as a consequence of the
functional  equation~\eqref{fe}  satisfied  by the  minimal  parabolic
Eisenstein series, the maximal parabolic Eisenstein series satisfy the
following  relationships, among  many  others:
\begin{eqnarray}
\bE^{E_{d+1}}_{[1\, 0^d];\frac32} &\propto&  \bE^{E_{d+1}}_{[0\,1\,0^{d-1}];1}\propto \bE^{E_{d+1}}_{[0^d\,1];{d-2\over2}} \,\nn\\
 \bE^{E_{d+1}}_{[1\, 0^d];\frac52} &\propto&   \bE^{E_{d+1}}_{[0^d\,1];{d+2\over2}}\,
 \label{equiveisen}
\end{eqnarray}
The symbol $\propto$ means that the
quantities are equal up to a constant of proportionality, which may depend on $d$.
In this manner our solutions can be rewritten in many different ways.  Such a relationship was pointed out in the $s=3/2$ case in~\cite{Obers:1999um}.

We will now check that the solutions in~\eqref{e:R4ans} and~\eqref{e:D4R4ans} behave in the appropriate manner in the three degeneration limits described in the introduction.

\bigskip
{\bf (i) Decompactification from $D$ to $D+1$}
\medskip

This is the limit associated with the parabolic subgroup $P_{\alpha_{d+1}}$, for node $d=10-D$.
Consistency under  decompactification in this limit  $r_d/\ell_{D+1}\gg1$  requires (see equations~(2.10) and~(2.11) of~\cite{Green:2010}),
\be
\label{e:DecLimitone}
  \int_{P_{\alpha_{d+1}}}    \cE^{(D)}_{(0,0)} \, \deq\,
{\ell_{D+1}^{8-D}\over \ell_{D}^{8-D}  }\,
\left({ r_{d}\over \ell_{D+1}  }\cE^{(D+1)}_{(0,0)}+\left(r_d\over\ell_{D+1}\right)^{8-D}\right)\,,
\ee
and
\be
\int_{P_{\alpha_{d+1}}} \cE^{(D)}_{(1,0)}\, \deq \,
{\ell_{D+1}^{12-D}\over\ell_{D}^{12-D}}  \,
\left({r_{d}\over\ell_{D+1}}\,\cE^{(D+1)}_{(1,0)}+\left(r_d\over\ell_{D+1}\right)^{6-D}\cE^{(D+1)}_{(0,0)}+\left(r_d\over\ell_{D+1}\right)^{12-D}\right)\, .\\
\label{e:DecLimittwo}
\ee
The symbol $\simeq$  means that constant factors multiplying each term have been suppressed.The fact that, for $D=3,4,5$ (i.e., for the duality groups $E_6$, $E_7$ and $E_8$)  the automorphic coefficients in these expressions are simply given by the Eisenstein series  shown in~\eqref{e:R4ans},~\eqref{e:D4R4ans} follows from the precise expressions given in the first three rows of tables~\ref{tab:Udual2}
and~\ref{tab:Udual3} (where the
parameter $r$ used in these expressions  is given by
$r^2=r_{d}/\ell_{D+1}$).

The     terms     linear     in     $r_d$     in~\eqref{e:DecLimitone}
and~\eqref{e:DecLimittwo} are  the ones that  contain the coefficients
of the interactions  in $D+1$ dimensions in the  large-$r_d$ limit, as
can      be     seen     by      substituting     in~\eqref{e:R4effec}
and~\eqref{e:D4R4effec}.    The   other    terms   contribute   in   a
well-understood  way  to  the  non-analytic  part  of  the  amplitude,
$A^{nonan}$ in~\eqref{ampsplit}  (or, equivalently, to  nonlocal terms
in the effective action) \cite{Green:2010wi}.  Schematically, the last
terms  in parentheses are  the $n=0$  and $n=2$  terms in  an infinite
series of  the form $r_d^{8-D}\sum_{n=0}^\infty  c_n\, (r_d^2\, s)^n\,
\cR^4$ that gives  the supergravity threshold behaviour $s^{(D-8)/2}\,
\log s\, \cR^4$ ($D$ even)  or $s^{(D-8)/2}\, \cR^4$ ($D$ odd), in the
limit $r_d^2s\to  \infty$ (this is the  standard one-loop supergravity
threshold).  This  can be described as  the effect of the  sum over an
infinite  set of thresholds  for Kaluza--Klein  modes of  mass $p/r_d$
(positive integer $p$) in the limit that they become massless.  The second term in parentheses in \eqref{e:DecLimittwo} is similarly the $n=0$ contribution to a second infinite series of powers, $r_d^{2-D}\sum_{n=0}^\infty c_n' (r_d^2 s)^{n+2}$, which sum up to give a threshold that arises at order $\ell_D^6$.  The structure of this term is determined by unitarity in the manner described in section 3 of \cite{Green:2006gt}.

Using the relationships derived in section~\ref{sec:eisenseries} it is straightforward to check that these properties are satisfied by the expressions~\eqref{e:R4ans} and~\eqref{e:D4R4ans}.
 In this manner it is clear that the chain of coefficients for dimensions $3\le D \le 10$ can be obtained from the functions for  $D=3$ (the $E_8$ case) in~\eqref{e:R4ans} and~\eqref{e:D4R4ans}. Furthermore, the expression~\eqref{e:DecLimittwo} shows that both the coefficients  $\calE^{(D+1)}_{(0,0)}$ and $\calE^{(D+1)}_{(1,0)}$ are contained in the constant term for $\calE^{(D)}_{(0,0)}$ in the parabolic subgroup $P_{\alpha_{d+1}}$.
In the next section we will see that the coefficient $\calE^{(D)}_{(0,1)}$ contains all three functions
$\calE^{(D+1)}_{(0,0)}$, $\calE^{(D+1)}_{(1,0)}$  and $\calE^{(D+1)}_{(0,1)}$ in its constant term in the $P_{\alpha_{d+1}}$ parabolic.  In particular, we will show that starting from the largest-rank case, $D=3$ with the group  $E_8$,  the complete towers of all three coefficients can be obtained by  starting from the single coefficient, $\calE^{(3)}_{(1,0)}$.

 Intriguingly,  line three of table~\ref{tab:Udual3}  shows that the coefficient $\calE^{(D)}_{(1,0)}$ demonstrates a bifurcation under the decompactification from $D=5$ to  $D=6$,  so that for $D= 6$ this coefficient is the sum of  two Eisenstein series (as suggested in~\cite{Green:2010wi}).  This point merits special discussion, which will be given in subsection~\ref{sec:decomfive}.

\bigskip
{\bf (ii)   The perturbative limit}
\medskip

This limit is associated with the parabolic subgroup $P_{\alpha_1}$ and is  given by $y_D\to 0$ with  $\ell_s$ fixed.  In this limit the expansions of
the interactions~(\ref{e:R4effec}) and~(\ref{e:D4R4effec}) are given by the constant terms~\cite{Green:2010wi},
\be
\label{e:Perlimitone}
\ell_{D}^{8-D}    \,\int_{P_{\alpha_1}}    \cE^{(D)}_{(0,0)} \,  \deq\,
\ell_{s}^{8-D}\,
\left({2\zeta(3)\over y_D}+ \bE^{SO(d,d)}_{[1\,0^{d-1}];{d\over2}-1}\right)\,,
\ee
and
\be
\ell_{D}^{12-D}    \,\int_{P_{\alpha_1}}    \cE^{(D)}_{(1,0)} \, \deq \,
\ell_{s}^{12-D}\,
\left({\zeta(5)\over  y_D}+  \bE^{SO(d,d)}_{[1\,0^{d-1}];{d\over2}+1}+
  y_D\, \bE^{SO(d,d)}_{[0^{d-2}\,10];2}\right)\,.
  \label{e:Perlimittwo}
\ee
Once again, the   precise  numerical coefficients  are   given  in   tables~\ref{tab:Udual2}
and~\ref{tab:Udual4} with the relation
$r^4=1/y_D$.

 A term of order $y_D^{-1+h}$ should be interpreted as a genus-$h$ perturbative string theory contribution to the amplitude that can be calculated by functional integration of a genus-$h$ world-sheet embedded in toroidally compactified Minkowski space, which reduces to an integral over the moduli of the world-sheet.  This can be checked explicitly by extending the methods of~\cite{Green:1999pv,Green:2008uj} to compactification on $\calT^d$ with $d>2$ and to $h\ge 1$.   This leads to expressions for  terms in the expansion of the genus-$h$ amplitude of the form
 $\sigma_2^p\sigma_3^q\,  \cR^4\,  I^{(d)}_h[j_h^{(p,q)}]$, where $j_h^{(p,q)}$ are
 functions of the world-sheet moduli that are, in principle,
 determined by expanding the genus-$h$ amplitude, although in practise
 this    has     only    been    done    in     detail    for    $h=1$
 in~\cite{Green:1999pv,Green:2008uj} and for the leading $(1,0)$ term at genus-two (the $\p^4 \, \cR^4$ term).

The leading terms in \eqref{e:Perlimitone} and \eqref{e:Perlimittwo}, the genus-zero parts ($h=0$) are
of order $y_D^{-1}$ and since  their coefficients do not depend on the
 compactification torus they are the same for all $D$.
The genus-one ($h=1$) amplitude is of order $y_D^0$ and  is given by the integral over the complex structure of the world-sheet torus, $\tau$,
of  the form
\begin{equation}
I^{(d)}_1[j_1^{(p,q)}]:= \int_{\cF_{SL(2,\mathbb Z)}} {d^2\tau\over\tau_2^2}\,
j_1^{(p,q)}(\tau)\,(\Gamma_{(d,d)}(\tau) -V_{d})\, ,
\end{equation}
where   $\cF_{SL(2,\mathbb   Z)}$   is   a  fundamental   domain   for
$SL(2,\mathbb Z)$ and $V_{d}=\sqrt{\det G}$ is the volume of $\calT^d$, and $G_{ij}$ is its metric: $G_{ij}=g_{ij}+b_{ij}$ with $g_{ij}$ a $d\times d$ symmetric
square matrix and $b_{ij}$ is a $d\times d$ antisymmetric square matrix.
.  The lattice factor is defined by
\begin{equation}
\Gamma_{(d,d)}(\tau) = V_{d}\sum_{(m^i,n^i)\in \mathbb Z^{2d}} e^{ -
  {\pi\over\tau_2} (m^i-n^i\tau)G_{ij}(m^j-n^j\bar\tau)}\,,
\label{lattfac}
\end{equation}
where repeated indices are summed.
Thus,  for the  $\cR^4$ term  with  coefficient $\calE^{(D)}_{(0,0)}$, since
$j^{(0,0)}_1 =1$ it follows that~\cite{Green:1999pv,Green:2008uj}
\begin{equation}
  \label{e:R4oneloop}
I^{(d)}_1[j^{(0,0)}_1] ={\Gamma({d\over2})\over\pi^{d\over2}} \bE^{SO(d,d)}_{[10\cdots 0];{d\over2}-1}\, ,
\end{equation}
which  matches the $y_D$-independent  term on  the right-hand  side of
\eqref{e:Perlimitone}.    The    $SO(d,d)$    series  will be  defined
in terms of explicit lattice sums in~(\ref{e:DnDef1})  (see appendix~C  of\cite{Green:2010wi}  for
details of  these series).

For  the  $\p^4\cR^4$  term  with  coefficient  $\calE^{(D)}_{(1,0)}$, since
$j^{(1,0)}_1=\bE^{SL(2)}_2(\tau)/(4\pi)^2$ it follows that~\cite{Green:1999pv,Green:2008uj}
\begin{equation}
  \label{e:D4R4oneloop}
I^{(d)}_1[j^{(1,0)}_1] ={2\zeta(4)\Gamma({d\over2}+1)\over8 \pi^{{d\over2}+1}}\, \bE^{SO(d,d)}_{[10\cdots 0];{d\over2}+1}\, .
\end{equation}

In the genus-two case, the functions $j_2^{(p,q)}$ that enter into the expression $I_2[j_2^{(p,q)}]$  have not been determined beyond the leading term, which is simply $j_2^{(1,0)}=1$.  This contributes to the  $\sigma_2\,\cR^4$ coefficient, $\calE^{(D)}_{(1,0)}$, which is given  by the integral  of the genus  two lattice
factor $\Gamma_{(d,d)}(\Omega)$
\begin{equation}
  \label{e:D4R4twoloopHD}
I^{(d)}_2[j^{(1,0)}_2] :=  \int_{\cF_{Sp(2,\mathbb
    Z)}}{|d^3\Omega|^2\over(\det \Im \Omega)^3}\,\Gamma_{(d,d)}(\Omega) \,,
\end{equation}
where $\Omega$ is the genus two period matrix and $\cF_{Sp(2,\mathbb
    Z)}$  is a  Siegel fundamental  domain for  $Sp(2,\mathbb  Z)$ and
  $\Gamma_{(d,d)}(\Omega)$ is the lattice sum defined by
  \begin{equation}
    \Gamma_{(d,d)}(\Omega)=(V_{d})^2\,       \sum_{(m^i_a,n^{ia})\in
      \mathbb    Z^{4d}}\,     e^{-\pi   G_{ij}
    (m^i_a-\Omega_{ab}n^{ib})(\Im\Omega^{-1})^{ac}(m^j_c-\Omega_{cd}n^{jd})}\,.
  \end{equation}
Using results from~\cite{Obers:1999um,Green:2010wi}, it is possible to deduce  from~\eqref{e:D4R4twoloopHD} that
\begin{equation}
  \label{e:D4R4twoloopHDI}
I^{(d)}_2[j^{(1,0)}_2] ={1\over 6\pi}\,(\bE^{SO(d,d)}_{[0\cdots 010];2}+\bE^{SO(d,d)}_{[0\cdots 01];2}\, ),
\end{equation}
for $d\leq 4$, and
\begin{equation}
  \label{e:D4R4twoloopHDII}
I^{(d)}_2[j^{(1,0)}_2] ={1\over 3\pi}\,\bE^{SO(d,d)}_{[0\cdots 01];2}\, ,
\end{equation}
for $d\geq5$.
In order to compare these perturbative string theory results with the constant term in \eqref{e:Perlimittwo} we need to use the  following relations between the $SO(d,d)$ series
\begin{eqnarray}
  \nn  \bE^{SO(d,d)}_{[0\cdots010];2}&=&\bE^{SO(d,d)}_{[0\cdots01];2}\\
    \bE^{SO(d,d)}_{[0\cdots010];d-3} &=&\bE^{SO(d,d)}_{[0\cdots01];d-3}\\
 \nn   \bE^{SO(d,d)}_{[0\cdots01];2}&\propto& \bE^{SO(d,d)}_{[0\cdots 10];d-3}\, .
\end{eqnarray}
The last equation is a direct consequence of the functional equations~\eqref{fe} for the $SO(d,d)$ Eisenstein series.
These string theory results are in accord with the term linear in $y_D$ in the constant term~\eqref{e:Perlimitone}.

\bigskip
{\bf (iii) Semi-classical M-theory}
\medskip

This   is   the  limit   associated   with   the  parabolic   subgroup
$P_{\alpha_2}$.  In this limit  the volume $\calV_{d+1} \to \infty$ of
the M-theory torus becomes large and the semi-classical, or Feynman diagram,
approximation to eleven-dimensional supergravity is useful.  The constant term of the coefficients in~(\ref{e:R4effec}) and~(\ref{e:D4R4effec})  in this parabolic subgroup is given by  (using the relation $\ell_D^{D-2}=\ell_{11}^9/\cV_{d+1}$, as well as $r^{1+d}=(\cV_{d+1}/\ell_{11}^{d+1})^{3/2}$ with $d=10-D$)~\cite{Green:1997as,Russo:1997mk,Green:1999pu,Green:2010wi},
\be
\label{e:MthLimitone}
\ell_{D}^{8-D}    \,\int_{P_{\alpha_2}}    \cE^{(D)}_{(0,0)}  \,\deq\,
{\cV_{d+1}\over\ell_{11}^3}\,
\left(4\zeta(2)+ \left(\ell_{11}^{d+1}\over\cV_{d+1}\right)^{3\over d+1}\bE^{SL(d+1)}_{[1\,0^{d-1}];{3\over2}}\right)\,,
\ee
and
\bea
\nn  \int_{P_{\alpha_d}}    \cE^{(D)}_{(1,0)} \,&\deq&\,
{\ell_{11}\,\cV_{d+1}\over\ell_{D}^{12-D} }  \,
\Big(\left(\cV_{d+1}\over\ell_{11}^{d+1}\right)^{1\over
    d+1}\bE^{SL(d+1)}_{[1\,0^{d-1}];-\frac12}
 +\left(\ell_{11}^{d+1}\over\cV_{d+1}\right)^{5\over d+1}
  \bE^{SL(d+1)}_{[1\,0^{d-1}];\frac52}\\
&+&\left(\ell_{11}^{d+1}\over \cV_{d+1}\right)^{8\over d+1}\bE^{SL(d+1)}_{[01\, 0^{d-2}];2}\Big)\,.
\label{e:MthLimittwo}
\eea
The    precise  values  of    the   constants    are    given    in
tables~\ref{tab:Udual2} and~\ref{tab:Udual5}.

The various contributions in~\eqref{e:MthLimitone}  agree with the expressions obtained by
evaluating the sum of one-loop and two-loop Feynman diagram contributions to the amplitude in eleven-dimensional supergravity  compactified  on a
$(d+1)$-torus~\cite{Green:1997as,Russo:1997mk,Green:1999pu}.  The two terms in  the $\cR^4$ coefficient
\eqref{e:MthLimitone}  arise from  the compactified  one-loop diagrams
together  with  the   counterterm  diagram as  in~\cite{Green:1997as},
while the terms in $\p^4\cR^4$ coefficient~\eqref{e:MthLimittwo} arise
from the  sum of the  compactified two-loop diagrams and  the one-loop
diagram    that     includes    a    vertex     for    the    one-loop
counterterm~\cite{Green:1999pu}.  We  refer  to appendices~B  and~G
of~\cite{Green:2010wi} for
the precise connection between these computations and the $SL(d+1)$ series entering
in~\eqref{e:MthLimittwo}.

\subsection{Bifurcation of $\calE^{(D)}_{(1,0)}$  between $D=5$ and  $D=6$}
\label{sec:decomfive}

The decompactification of the $D=5$ coefficient $\calE^{(5)}_{(1,0)}$,
\begin{equation}
  \label{e:D4R4D5}
  \cE^{(5)}_{(1,0)}={1\over2}\,\bE^{E_6}_{[100000];\frac52}\, ,
\end{equation}
merits special discussion.
As noted earlier, the third line of table~\ref{tab:Udual3} shows that the constant term of the $E_6$ Eisenstein series, corresponding to the decompactification from $D = 5$ to $D=6$ results in the sum of  two $SO(5,5)$ series in the combination
${1\over 2} \hbE^{SO(5,5)}_{[10000];\fiveh}+ {4\over 45}\, \hbE^{SO(5,5)}_{[00001];3}$ mutiplying $r^{20/3}$,
where  the hats indicate   the finite  part  of the  series
after   subtraction     of   an   $\epsilon$     pole    as
in~\cite{Green:2010wi}.
Although the individual $SO(5,5)$ series  have poles in $s$, the residues of these poles cancel and the sum is finite (as discussed in~\cite{Green:2010wi,Green:2010}).

This is seen by using the   relations  $\ell_5^3=   \ell_6^4  /r_5$  and
$r=(r_5/\ell_6)^{1/2}$, leading to
\begin{equation}\begin{split}
  \label{e:D4R4D5dec}
 \int_{P_{\alpha_6}} \, \cE^{(5)}_{(1,0)}&={\ell_6^6 r_5\over\ell_5^7}\,
  \left({1\over                                                      2}
    \hbE^{SO(5,5)}_{[10000];\frac52}+{4\over45}\,\hbE^{SO(5,5)}_{[00001];3}\right.\cr
    &\left. \qquad\ \
  +2\log\left(r_5\over\ell_6\mu\right)\,\cE^{(6)}_{(0,0)}+ {\zeta(7)\over 6} \,\left(r_5\over\ell_6\right)^6 \right)\,.
\end{split}\end{equation}
where $\mu$ is a constant scale factor.
The term linear in $r_5$ is the one that multiplies the $D=6$ coefficient, $\calE^{(6)}_{(1,0)}$ so that
\begin{eqnarray}\label{e:D4R4D6def}
 \cE^{(6)}_{(1,0)}=\frac12\, \hbE^{SO(5,5)}_{[10000];\frac52}+{4\over45}\,\hbE^{SO(5,5)}_{[00001];3}\, .
\end{eqnarray}
In order to interpret the $\log r_5$ term in \eqref{e:D4R4D5dec} we need to analyze properties of the $SO(5,5)$ series in this expression.
 Although such properties are  contained in the general expressions in
 section~\ref{sec:eisenseries}, it is illuminating to obtain them from
 the representation of  such series as lattice sums.   In general such
 lattice sums are  awkward to analyze but in  this case the expression
 for       $\bE^{SO(d,d)}_{[1\,0^{d-1}];s}$       (given      in~(C.2)
 of~\cite{Green:2010wi})  is  expressible  in  a useful  manner.   The
 expression given in~\cite{Green:2010wi} is a provided by a Siegel-Weil
 formula~\cite{Weil} relating the integral  over the
moduli space  of genus-one Riemann surfaces  of $SL(2,\IZ)$ Eisenstein
series times lattice sums and $SO(d,d)$ Eisenstein series,
 \begin{equation}\label{e:DnDef1}
\bE^{SO(d,d)}_{[1\,0^{d-1}];s}=   {\pi^{s}\over 2\zeta(2s+2-d)\Gamma(s)}\,\int_{\calF_{SL(2,\mathbb Z)}} {d^2\tau\over
     \tau_2^2} \bE_{s+1-{d\over2}}(\tau)\,(\Gamma_{(d,d)}(\tau)-V_{d})\,,
\end{equation}
where $\bE_s(\tau)=  \sum_{(m,n)\neq(0,0)} \, y^s/|m+n\tau|^{2s}$
is    the   usual    $SL(2,\mathbb   Z)$    Eisenstein    series   and
$\Gamma_{(d,d)}(\tau)$ is defined in~\eqref{lattfac}.
It follows from this definition that the series satisfies the
functional equation
\begin{equation}
  \label{e:FunSOdd}
  \bE^{SO(d,d)}_{[1\,0^{d-1}];s}          ={\xi(2s-2d+3)\xi(2s-d+1)\over
    \xi(2s)\xi(2s-d+2)}\, {\zeta(2s)\over\zeta(2d-2-2s)}\,\bE^{SO(d,d)}_{[1\,0^{d-1}];d-1-s}\, ,
\end{equation}
where  $\xi(s)$ is the  completed Riemann  $\zeta $-function  $\xi (s)
=\pi^{-s/2}\Gamma({s\over  2})  \zeta(s)$.  This  functional  relation
implies that the $SO(d,d)$ series has a single pole at $s=d/2$
\begin{equation}
  \label{e:SOpole}
  \bE^{SO(d,d)}_{[1\,0^{d-1}];\frac{d}2+\epsilon}=\f{6}{d-2}
  {\bE^{SO(d,d)}_{[1\,0^{d-1}];\frac{d}2-1}\over                      \epsilon}+
  \hbE^{SO(d,d)}_{[1\,0^{d-1}];\frac{d}2}+O(\epsilon)\, .
\end{equation}
The  residue of  the series  $\bE^{SO(d,d)}_{[0\cdots01];s}$ at
$s=(d+1)/2$ can be extracted using the methods of  section~\ref{sec:eisenseries}.

For the $SO(5,5)$ case this becomes
\begin{equation}
  \label{e:ESO55V}
 \bE^{SO(5,5)}_{[10000];\frac52+\epsilon}=
  {2\bE^{SO(5,5)}_{[10000];\frac32}\over                      \epsilon}+
  \hbE^{SO(5,5)}_{[10000];\frac52}+O(\epsilon)\, .
\end{equation}
The  general methods  of section~\ref{sec:eisenseries}  indicates that
the series $\bE^{SO(5,5)}_{[00001];s}$ has a single pole at $s=3$ given by
\begin{equation}
  \label{e:ESO55C}
 \bE^{SO(5,5)}_{[00001];3+\epsilon}={45\over 4}\,
  {\bE^{SO(5,5)}_{[10000];\frac32}\over                      \epsilon}+
  \hbE^{SO(5,5)}_{[00001];3}+O(\epsilon)\,.
\end{equation}
It is striking that the residue of the pole is given by the $\cR^4$ coefficient,
$\cE^{(6)}_{(0,0)}=\bE^{SO(5,5)}_{[10000];3/2}$.  This  is the  reason
why     the     coefficient     of     the     $r_5\,\log(r_5/\ell_6)$ term
in~\eqref{e:D4R4D5dec} is  the $\cE^{(6)}_{(0,0)}$
coefficient (see equation~(2.11) of~\cite{Green:2010} with $D=5$).

Because  the  residues  of  the $\epsilon$  poles  in~\eqref{e:ESO55V}
and~\eqref{e:ESO55C} are both proportional to $\bE^{SO(5,5)}_{[10000];\frac32}$ the $\p^4 \cR^4$ coefficient in $D=6$ can be written as the limit
\begin{equation}
    \cE^{(6)}_{(1,0)}= {1\over 2}\,\lim_{\epsilon\to0}
     \left(\bE^{SO(5,5)}_{[10000];\frac52+\epsilon}+{8\over45}\,\bE^{SO(5,5)}_{[00001];3-\epsilon}\right)
\end{equation}
as suggested in~\cite{Green:2010wi}.
 The cancellation of divergences is a general feature of all coefficient functions for all values of $D$ and is consistent with the absence of ultraviolet divergences in string theory.

  The coefficient $\calE^{(D)}_{(1,0)}$ for $5< D <10$ continues to be
  given by  the sum of two distinct  series~\cite{Green:2010wi} and it
  is only for the exceptional groups and $SL(2,\IZ)$ that this coefficient is given by a single Eisenstein series.

\section{Constant terms for the solution of the $E_8$ inhomogeneous Laplace eigenvalue equation}
\label{sec:inhomogeneous}
The previous sections concern the coefficients in the effective action that satisfy the Laplace eigenvalue equations~\eqref{laplaceeigenone} and~\eqref{laplaceeigentwo} with solutions that are Eisenstein series, whereas
the coefficient $\calE^{(D)}_{(0,1)}$ of the $\partial^6 \cR^4$ coefficient in~\eqref{e:D6R4effec} satisfies the inhomogeneous equation~\eqref{laplaceeigenthree}.
Its solution, subject to appropriate boundary conditions,   is formally given in terms of a Green  function by (for $D\ne 6$)
\be
\cE^{(D)}_{(0,1)}=\cE^{(D)}_{hom}- \left( \Delta^{(D)} -{6(14-D) (D-6)\over D-2} \right)^{-1}\, \left(\cE_{(0,0)}^{(D)}\right)^2 \,,
\label{solutionthree}
\ee
 where  $\cE^{(D)}_{hom}$  is   a  solution   to  the
homogeneous equation.
 Expressing this solution in a more explicit manner is a challenge
which we will not undertake here (see~\cite{Green:2005ba} for a discussion of the   $SL(2,\IZ)$ case).
Instead,  we  will  study  the  appropriate  constant  terms  of  this
coefficient in two of the parabolic subgroups, $P_{\alpha_1}$ and
$P_{\alpha_{d+1}}$, of relevance in this paper, as was done for the cases $6\le D\le 10$ (i.e., for duality groups of rank $r\le 5$) in
\cite{Green:2005ba,Basu:2007ck,Green:2010wi,Green:2010}.
 We shall omit
the  analysis of  the  third subgroup  $P_{\alpha_2}$  for economy  of
space,  since  it  presents few  new  issues.  Furthermore, for the sake of brevity,
rather than considering all the remaining cases ($D=3,4,5$),
we will consider the fundamental example of the $D=3$ coefficient, $\calE^{(3)}_{(0,1)}$, from which the others can be obtained.
This is an   $E_8$ automorphic function
that satisfies the inhomogeneous Laplace eigenvalue equation~\eqref{laplaceeigenthree},
\begin{equation}\label{e:D6R43D}
  (\Delta^{(3)}+198) \, \cE^{(3)}_{(0,1)}= - (\cE^{(3)}_{(0,0)})^2\, ,
\end{equation}
where the source term involves the square of
\begin{equation}
    \cE^{(3)}_{(0,0)}= \bE^{E_8}_{[10000000];\frac32}\, .
\end{equation}
  As  we will  see,  the
requirement that these constant  terms have the correct structure once
again determines them.

 We note that  the issue of whether the  solution \eqref{solutionthree} is unique
depends on  whether there  is a solution  to the  homogeneous equation with vanishing constant terms in the parabolic subgroups $P_{\alpha_{d+1}}$ and $P_{\alpha_1}$ (limits~(i),~(ii))    considered in
this section.  Such a solution would be an automorphic
function with eigenvalue $-198$.  We have verified that that no maximal
parabolic  Eisenstein  series  satisfies  the boundary  conditions  in
either of the two limits  under consideration, but it is possible (but
seems unlikely) that some more general series with the same eigenvalue
may satisfy them. There are also potentially cusp form solutions to
  the  homogeneous  equation,  though  these also seem  unlikely  to  exist
  because of the reasons mentioned in the introduction.   Their presence  would   amount  to  an  ambiguity  in  the
solution.

\bigskip
\noindent{\bf (i) Decompactification to $D=4$}\medskip

Before specialising to $D=3$ (the $E_8$ case) we will review the behaviour of $\calE^{(D)}_{(0,1)}$  for general $D=10-d$ in this limit.   Based on input from string theory the constant term in the parabolic subgroup $P_{\alpha_{d+1}}$, the decompactification limit, should  consist of five components with distinct powers of $r$  (see equation~(2.12) of~\cite{Green:2010}),
\begin{equation}\begin{split}
\int_{P_{\alpha_{d+1}}}&\cE^{(D)}_{(0,1)}\, \deq\, \left(\ell_{D+1}\over\ell_{D}\right)^{14-D}\,\Big( {r_d\over\ell_{D+1}}\,
\cE^{(D+1)}_{(0,1)}       +  \left(r_d\over    \ell_{D+1}\right)^{14-D} \cr
&
+   \left({r_d\over    \ell_{D+1}}\right)^{8-D}
\cE^{(D+1)}_{(0,0)}      +     \left({r_d\over    \ell_{D+1}}\right)^{4-D}
\cE^{(D+1)}_{(1,0)} \cr
& +\left(r_d\over    \ell_{D+1}\right)^{15-2D}+O(e^{-r_d/\ell_{D+1}}) \Big)\,
\label{d6r4series}
\end{split}\end{equation}
(certain $\log r$ factors that arise for specific values of $D$ have been suppressed in this expression).  The following is a sketchy summary of the interpretation of the five components of this expression.  The term proportional to $r_d$  contains the coefficient $\cE^{(D+1)}_{(1,0)} $  in the decompactified theory.  The second term, which has a constant coefficient, is the $n=3$ contribution to the infinite series of $(s r_d^2)^n\, \cR^4$ terms ($n\ge 0$) that generates the one-loop supergravity threshold behaviour in $D+1$ dimensions (the first two terms arose in \eqref{e:DecLimitone} and \eqref{e:DecLimittwo}).   Similarly, the third term, proportional to $\calE^{(D+1)}_{(0,0)}$, is the $n=1$ term in a series of terms of the form  $r_d^{2-D}(s r_d^2)^{n+2} \, \cR^4$ that contributes to the second threshold in $D+1$ dimensions (the $n=0$ term arose in~\eqref{e:DecLimittwo}).  The fourth term, proportional to $\calE^{(D+1)}_{(1,0)}$,  is the $n=0$ term in a new infinite series of terms of the form $r_d^{-(2+D)}(s r_d^2)^{n+3}\,\cR^4$ that generates the correct $(D+1)$-dimensional behaviour of a new threshold contribution at order $\ell_D^8$.  Finally, the last term in parentheses  is the $n=0$ term in a series of terms of the form $r_d^{-2D+9}(s r_d^2)^{n+3}\,\cR^4$, which sums to the two-loop supergravity threshold in $D+1$ dimensions.   Again the structure of these terms that contribute to thresholds is in accord with unitarity, generalising the discussion in \cite{Green:2006gt}.

 Thus, as stated earlier, the constant term of the series $\calE^{(D)}_{(0,1)}$ in this parabolic subgroup contains the information concerning all three coefficients  in $D+1$ dimensions.

We now consider the constant term that arises from the solution of  \eqref{e:D6R43D}, which describes limit (i) in
the $E_8$ case. We are interested in the limit   associated   with   the  parabolic   subgroup
$P_{\alpha_8}$, associated with the right-hand node of the $E_8$ Dynkin diagram of fig.~\ref{fig:dynkin}.
As a template for what this constant term should look like, we note that the constant term of the source term can be expressed as
\begin{equation}\begin{split}
      \int_{P_{\alpha_8}}         (\cE^{(3)}_{(0,0)})^2&=
{\ell_4^{11}\over\ell_3^{11}}\,   \Big({r_7\over\ell_4}\,(\bE^{E_7}_{[1\,0^6];\frac32})^2           +\left(r_7\over\ell_4\right)^5\,{6\zeta(5)\over\pi}\,
 \bE^{E_7}_{[1\,0^6];\frac32} \cr
&+\left(r_7\over\ell_4\right)^{9} \, {9\zeta(5)^2\over\pi^2}+O(e^{-r_7/\ell_4})\Big)\,.
\label{const3D}
\end{split}
\end{equation}
Based on the structure of this expression, together with the form anticipated in \eqref{d6r4series} with $D=3$, we make the ansatz
\begin{equation}\begin{split}
    \int_{P_{\alpha_8}}         \cE^{(3)}_{(0,1)}&=
{\ell_4^{11}\over\ell_3^{11}}\,\Big({r_7\over\ell_4} F_1^{E_7}+ {r_7\over\ell_4}\log \left( { r_7\over\ell_4\mu}\right) \,
  F_2^{E_7}                    +\left(r_7\over\ell_4\right)^5\,
  F_3^{E_7}\cr
&+\left(r_7\over\ell_4\right)^{9}\,F^{E_7}_4+\left(r_7\over\ell_4\right)^{11}\,F^{E_7}_5+O(e^{-{r_7/\ell_4}}) \Big)\,,
\end{split}\label{d6r4seriestwo}\end{equation}
where $\mu$ is a constant scale factor in the logarithm (the presence of which will become clear later) and the coefficients $F_r^{E_7}$ are coefficients that will now be determined.

In the limit corresponding to this parabolic the Laplacian
$\Delta^{(3)}$   on  $E_8/SO(16)$  decomposes into  a   sum  of   the  Laplacian
$\Delta^{(4)}$ on $E_7/SU(8)$ and the Laplacian along the $r_7$ direction (as in appendix~H.2
of~\cite{Green:2010wi}),
\begin{equation}\label{e:D8toD7}
  \Delta^{(3)}\to \Delta^{(4)}+\frac14\, (r_7\partial_{r_7})^2-{29\over2}\,r_7\partial_{r_7}\,.
\end{equation}
Substituting this expression together with~\eqref{const3D} and~\eqref{d6r4seriestwo} into~\eqref{e:D6R43D}  (and using $\ell_3=\ell_4^2/r_7$)  suggests matching terms as follows:
\bea
  (\Delta^{(4)}+60)\,          F^{E_7}_1&=&         -
(\bE^{E_7}_{[1\,0^6];\frac32})^2\,,
\label{e:Fn1}\\
 (\Delta^{(4)}+60)\, F^{E_7}_2&=& 0\,,
\label{e:Fn2}\\
(\Delta^{(4)}+30)\,      F^{E_7}_3&=&     -{6\zeta(5)\over\pi}
 \bE^{E_7}_{[1\,0^6];\frac32}  \,,
\label{e:Fn3}\\
(\Delta^{(4)}+8)\,      F^{E_7}_4&=&   - {9\zeta(5)^2\over\pi^2} \,,
\label{e:Fn4}\\
\Delta^{(4)}\, F^{E_7}_5&=&0\,.
\label{e:Fn5}
\eea
In deriving these equations the relations
\begin{equation}
 (\frac14\, (r_7\partial_{r_7})^2-{29\over2}\,r_7\partial_{r_7}+198)\, r_7^{22}=0\,,
\end{equation}
and
\begin{equation}
 (\frac14\, (r_7\partial_{r_7})^2-{29\over2}\,r_7\partial_{r_7}+138)\,
 r_7^{12}\log(r_7)=- {17\over2}\,r_7^{12}\,,
\end{equation}
were used, which accounts for the $\log r_7$ factor in  \eqref{d6r4seriestwo}.
Equation~\eqref{e:Fn1} is the equation satisfied by  the $\p^6\,\cR^4$  coefficient in  four
dimensions (\eqref{laplaceeigenthree} with $D=4$), so we identify
\begin{equation}
  \label{e:F1}
   F_1^{E_7}= \cE^{(4)}_{(0,1)}\, .
\end{equation}
This uses the fact that the right-hand side of~\eqref{e:Fn1} involves the square of the $D=4$ $\cR^4$ coefficient,
\begin{equation}
  \cE^{(4)}_{(0,0)}= \bE^{E_7}_{[1\,0^6];\frac32}\, .
\end{equation}
In principle, we could repeat this procedure and study the constant term of the $E_7$ coefficient in the parabolic subgroup $P(\alpha_7)$, and so on, in order to match with the known coefficients for the lower-rank cases.
Equation~\eqref{e:Fn2} is solved by the maximal parabolic Eisenstein series  that
is proportional to the $\p^4\,\cR^4$ coefficient in four dimensions,
\begin{equation}\label{e:F2}
  F_2^{E_7} \ \propto \  \bE^{E_7}_{[1\, 0^6];\frac52}  \ \propto \ \calE^{(4)}_{(1,0)}\ ,
\end{equation}
as required
by the general formula~\eqref{d6r4series} for $D=3$.  The discussion of the $\partial^4\cR^4$ interaction in section~\ref{sec:detaildegen} demonstrates that this satisfies the appropriate boundary conditions.
Equation~\eqref{e:Fn3} is solved by
\begin{equation}
  \label{e:F3}
 F_3^{E_7}={\zeta(5)\over2\pi}\,\bE^{E_7}_{[1\, 0^6];\frac32}\,,
\end{equation}
which is proportional to $\calE^{(4)}_{(0,0)}$ given in \eqref{e:R4ans}, as it should be according to~\eqref{d6r4series}.
In the case of~\eqref{e:Fn4} none of the solutions of the homogenous equation
\begin{equation}
  (\Delta^{(4)}+8)\, F_4^{E_7}=0\, ,
\end{equation}
are  compatible  with  the   boundary  conditions  imposed  by  string
perturbation. Therefore the solution to~\eqref{e:Fn4} is the constant function
\begin{equation}\label{e:F4}
  F_4^{E_7}= -{9\zeta(5)^2\over 8\pi^2}\, ,
\end{equation}
Extracting the value of $F_5^{E_7}$ , which  satisfies~\eqref{e:Fn5}, is more subtle since the constant term expansion of the  source term in~(\ref{const3D})  does not include an explicit power of $r^{11}$.  Nevertheless,  the  full   solution,     $(\Delta^{(3)}+198)^{-1}\,
(\cE^{(3)}_{(0,0)})^2$, does contain such a contribution in  its zero mode
expansion  that can  be  extracted  by projecting  this solution onto  a
solution of the homogeneous equation, as in section~5.4 of~\cite{Green:2005ba}.
This gives rise to a constant solution,
\begin{equation}\label{e:F5}
 F_5^{E_7}=\textrm{const}\,.
\end{equation}
Although the values of the constants in~(\ref{e:F2}) and~(\ref{e:F5}) can be extracted from the complete solution using an extension of the methods in~\cite{Green:2005ba}, this will
not be carried out here.

In summary, we have determined the functions $F_1^{E_7},\dots ,F_5^{E_7}$ that are the power-behaved components of the constant term $\int_{P_{\alpha_8}} \calE^{(3)}_{(0,1)}$  defined by \eqref{d6r4seriestwo}, and this
matches the form \eqref{d6r4series} anticipated from string theory considerations.  Strikingly, this  constant term contains all three coefficients for the $E_7$ case, $\calE^{(4)}_{(0,0)}$, $\calE^{(4)}_{(1,0)}$, $\calE^{(4)}_{(0,1)}$, as is evident from~\eqref{d6r4series} with $D=3$.   Therefore all the preceding coefficients of the lower rank examples are contained in this one example.

\bigskip
\noindent{\bf (ii) The perturbative expansion}\medskip

The perturbative  expansion is given  by the constant  term associated
with  the maximal parabolic  subgroup associated with  the
node $P_{\alpha_1}$ with Levi subgroup $GL(1)\times SO(7,7)$.
String perturbation theory (an expansion in powers of $y_3$ starting at $y_3^{-1}$) requires this to have the form
\begin{equation}\label{pertcoeffs}
\ell_3^{11}  \int_{P_{\alpha_1}}\,    \cE^{(3)}_{(0,1)}=   \ell_s^{11}\,\left(
  \sum_{k=0}^3 {y_3}^{k-1}\,F_k^{SO(7,7)}+ O(e^{-1/y_3})\right)\,.
\end{equation}

The coefficients $F_k^{SO(7,7)}$ can be determined by a procedure analogous to the one in the previous limit, as follows.
First the Laplacian $\Delta^{(3)}$ on $E_8/SO(16)$ is decomposed in this
limit into a sum of the Laplacian on $SO(7,7)/(SO(7)\times SO(7))$ and a
Laplacian along  the $y_3$ direction\footnote{Requiring the tree-level contributions to $\cR^4$
  and    $\p^4\,\cR^4$  to be annihilated    by   the
  $SO(d,d)/(SO(d)\times SO(d))$ Laplacian implies
\begin{equation}
  \Delta^{(D)}  \to  \Delta^{SO(d,d)/(SO(d)\times SO(d))}+\frac{D-2}2
  (y_D\partial_{y_D})^2+{D^2-19D+94\over2}\,y_D\partial_{y_D}\, .\nn
\end{equation}}
\begin{equation}
  \Delta^{(3)}  \to  \Delta^{SO(7,7)/(SO(7)\times SO(7))}+\frac12
  (y_3\partial_{y_3})^2+23\,y_3\partial_{y_3}\, .
\label{agara}
\end{equation}
Next,  the constant term of the source is obtained by substituting the expansion of
$\cE^{(3)}_{(0,0)}$ (given in table~\ref{tab:Udual2} with $r^{-4}=y_3$), resulting in
\begin{equation}\label{sourceexpand}
  \int_{P_{\alpha_1}}\,                          (\cE^{(3)}_{(0,0)})^2=
  {\ell_s^{11}\over\ell_3^{11}}\, \left({4\zeta(3)^2\over y_3}+{6\over
    \pi}\,\bE^{SO(7,7)}_{[1\,0^6];\frac32}+ {9y_3\over4\pi^2}\, (\bE^{SO(7,7)}_{[1\,0^6];\frac32})^2+O(e^{-1/y_3})\right)\, .
\end{equation}
The structure of this expression is consistent with \eqref{pertcoeffs}, which we may use as an ansatz for the solution.
Substituting~\eqref{pertcoeffs}, (\ref{agara}) and~\eqref{sourceexpand}  into (\ref{e:D6R43D}) results in equations that determine the coefficients $F_k$ (using $\ell_3=\ell_s y_3$),
\bea
\label{genusdos0}
(\Delta^{SO(7,7)/SO(7)\times SO(7)} -6)\, F_0^{SO(7,7)} &=& -4\zeta(3)^2\,,\\
(\Delta^{SO(7,7)/SO(7)\times    SO(7)}+{11\over2})\,    F_1^{SO(7,7)}&=&   -
    {6\zeta(3)\over\pi}\, \bE^{SO(7,7)}_{[1\,0^6];\frac52}\, ,
\label{genusdos1}\\
(\Delta^{SO(7,7)/SO(7)\times    SO(7)}+18)\,    F_2^{SO(7,7)}&=&   -
    {9\over4\pi^2}\, (\bE^{SO(7,7)}_{[1\,0^6];\frac52})^2\, ,
\label{genusdos2}\\
(\Delta^{SO(7,7)/SO(7)\times    SO(7)}+{63\over2})\,    F_3^{SO(7,7)}&=& 0\ .
\label{genusdos3}
\eea
A solution to \eqref{genusdos0} that is compatible with string perturbation theory is the constant
\begin{equation}
  F_0^{SO(7,7)}={2\zeta(3)\over3}\, ,
\end{equation}
which is precisely the genus zero (tree-level) contribution~\cite{Green:2010wi}.
A solution to the homogeneous equation (\eqref{genusdos1} with no source term)
that is  consistent with string theory is $\bE^{SO(7,7)}_{[1\,0^{6}];{11\over2}}$, resulting in a
solution of~\eqref{genusdos1}  given by
\begin{equation}
  F^{SO(7,7)}_1=    \frac{1}{12}\,    \bE^{SO(7,7)}_{[1\,0^{6}];{11\over2}}+
  {\zeta(3)\over2\pi}\, \bE^{SO(7,7)}_{[1\,0^{6}];{5\over2}}\,.
\end{equation}
This agrees with the genus one  string theory expression $  I^{(d)}_1[j^{(0,1)}_1]$ evaluated in
 appendix~D of~\cite{Green:2010wi} using $j_1^{(0,1)}=10 \bE_3(\tau)/(4\pi)^3 +\zeta(3)/32$.  The function $F_2^{SO(7,7)}$ satisfies the inhomogeneous Laplace eigenvalue equation (\ref{genusdos2}), which may be analysed in the same manner as analogous  examples considered in~\cite{Green:2005ba,Basu:2007ck,Green:2010wi,Green:2010}.
The  function, $F_3^{SO(7,7)}$,  satisfies the source-free (homogeneous) equation~\eqref{genusdos3}  since there is no $y_3^2$ term in the constant term of the source.  As we will see, a solution of relevance to string theory is  given by the linear combination of maximal parabolic Eisenstein series,
\begin{equation}\label{pertex}
    F_3^{SO(7,7)}=     \alpha\,     \bE^{SO(7,7)}_{[0^6\,1];3}+\beta\,
    \bE^{SO(7,7)}_{[0^5\,10];3}               +               \gamma\,
    \bE^{SO(7,7)}_{[0^2\,1\, 0^4];\frac32}\, ,
\end{equation}
where  $\alpha$,  $\beta$ and $\gamma $ are  constants that are determined  from the
boundary conditions.
By  a   direct  evaluation  of   the  series  using  the   methods  of
section~\ref{sec:eisenseries}  we find  that  these Eisenstein  series
satisfy the relations
\begin{eqnarray}
\nn    \bE^{SO(7,7)}_{[0000001];3}&=& \bE^{SO(7,7)}_{[0000010];3} \ ,\\
\bE^{SO(7,7)}_{[0010000];\frac32}&=&0\ .
\end{eqnarray}
Therefore the expression~(\ref{pertex}) takes the form
\begin{equation}\label{pertex2}
    F_3^{SO(7,7)}= (\alpha+\beta)\, \bE^{SO(7,7)}_{[0^6\,1];3}\ .
\end{equation}
The normalisation is fixed by comparison with the genus three contribution in  string theory in the limit in which the volume of the  $7$-torus, $\calT^7$,  is large  (see appendix~F  of~\cite{Green:2010wi}), resulting in
\begin{equation}
    \alpha+\beta= {\textrm{vol}(\cF_{Sp(3,\mathbb Z)})\over2\zeta(6)} =\frac{1}{270}\ ,
\end{equation}
where $\textrm{vol}(\cF_{Sp(3,\mathbb Z)})=\zeta(6)/135$ is the  volume of the Siegel fundamental domain for
$Sp(3,\mathbb Z)$~\cite{Siegel}.

To summarise, we have determined the constant terms  of the solution of equation~\eqref{e:D6R43D} for $\calE^{(3)}_{(0,1)}$  in the parabolic subgroup $P_{\alpha_1}$ that agree with the results of the explicit evaluation of string perturbation theory.

\section{Summary and comments}

In this paper  we have determined the expressions for  the coefficients  $\cE^{(D)}_{(0,0)}$,     $\cE^{(D)}_{(1,0)}$  of  the $\cR^4$, $\partial^4\, \cR^4$ interactions, the first two higher derivative terms in the low energy expansion of the four-supergraviton scattering amplitude in maximally supersymmetric string theory.   These coefficients are the maximal parabolic Eisenstein series in~\eqref{e:R4ans} and~\eqref{e:D4R4ans} for the duality groups $E_6(\IZ)$, $E_7(\IZ)$ and $E_8(\IZ)$.  All the lower rank cases that were determined in earlier work follow from these by considering the behaviour in the vicinity of the  decompactification cusp, limit~(i), defined by the root $\alpha_{d+1}$.    Indeed,  starting from the highest-rank case, where the duality group is $E_8(\IZ)$ and $D=3$, all the lower-rank cases are contained in the constant term for the parabolic subgroup defined by the root $\alpha_8$.  Thus, the constant term in limit~(i) leads to a chain  of nested finite combinations  of maximal parabolic Eisenstein series.   In other words, the coefficients  of the interactions for the lower rank duality groups can be deduced by successive decompactifications to space-time dimensions  $4\leq  D\leq 10$.
Although there might be ambiguities due to the presence of cusp forms, as discussed in the introduction, it is likely that these are absent for the kinds of series we are considering here.
 A curious feature arises on decompactifying  the five dimensional
$\p^4\,\cR^4$   interaction to   six  dimensions, where a bifurcation arises in which $\cE^{(5)}_{(1,0)}$ splits into the sum of two $SO(5,5)$ Eisenstein series in the $D=6$ theory.  Whereas each of these series has a pole at the value $s=5/2$, the poles cancel and the sum of the terms is analytic.  The  same phenomenon has previously been noted in dimensions $D\geq7$ in~\cite{Green:2010wi}.

In section~\ref{sec:inhomogeneous} we considered the coefficient,   of the $\partial^6\, \cR^4$ interaction in the $E_8$ case, $\calE^{(3)}_{(0,1)}$, which solves the inhomogeneous Laplace eigenvalue equation,~\eqref{e:D6R43D}, and is not an Eisenstein series.  Although we did not discuss the full solution in this case,  the constant term for this coefficient in the parabolic subgroup labelled by the root $\alpha_8$ was shown to contain all three of the $E_7$ automorphic coefficient functions  $\cE^{(4)}_{(0,0)}$,     $\cE^{(4)}_{(1,0)}$  and $\cE^{(4)}_{(0,1)}$, and therefore contains within it the complete chain of coefficients in dimensions $4\le D \le 10$, including those determined in earlier work.

Furthermore, we determined the  constant terms of these coefficient functions in the parabolic subgroup defined by the root $\alpha_1$ that  determines the behaviour in limit~(ii), the limit of string perturbation theory.  In almost all cases these power-behaved terms match  the  results obtained directly from type II superstring perturbation theory evaluated  on higher-dimensional tori, the only exceptions being those cases in which the string calculations have not yet been carried out.  Similarly the behaviour of the  $\cR^4$ and $\partial^4 \cR^4$ interactions in limit (iii),  in which the volume of the $(d+1)$-dimensional M-theory torus is large, is contained in the constant terms for the parabolic subgroup defined by the root $\alpha_2$.  These constant terms precisely match calculations in semiclassical eleven-dimensional supergravity based on one and two-loop Feynman diagrams~\cite{Green:1997as,Russo:1997mk,Green:1999pu,Green:2008bf} (the constant term of the $\partial^6\cR^4$ coefficient in this parabolic subgroup has yet to be considered).

These considerations made extensive use of the  properties of constant terms in order to match the  boundary data obtained from string theory and supergravity.   In the case of
Eisenstein  series  the constant terms  are  power  series in  the  parameter  $r$,
as defined in section~\ref{sec:description},
though the automorphic  function $\calE^{(D)}_{(0,1)}$ also contains
terms  that  are exponentially  small  in  the  large-$r$ limit (interpreted as instanton anti-instanton pairs with zero net instanton charge).   The
non-constant terms  of the functions we have discussed are also of great  interest within string  theory since
they are interpreted as  sums over contributions of charged instantons that are associated with euclidean
$p$-branes  of various  kinds (where  $0\le p  \le 6$)  wrapped around
$(p+1)$-cycles of the torus,  $\calT^d$. The spectrum of such instantons determines the behaviour of the non-zero Fourier components of the coefficient functions at the cusps.  Furthermore, the
instanton spectrum  in $D$ dimensions ($D=10-d$) is related to the spectrum of BPS particle states in $D+1$  dimensions, which are charged black holes of various kinds that play an important r{\^o}le in the higher-dimensional theory.  The  fractional  BPS    nature of the $\cR^4$, $\partial^4\, \cR^4$ and $\partial^6\, \cR^4$
 interactions is encoded  in the instanton
 measure  factors that enter the expansion of the coefficients $\calE^{(D)}_{(0,0)}$,
$\calE^{(D)}_{(1,0)}$ and $\calE^{(D)}_{(0,1)}$ near each cusp~\cite{Green:1998yf,Moore:1998et, Sugino:2001iq}.
  The  $1/8$-BPS configurations  that  enters in
$\calE^{(D)}_{(0,1)}$   are  particularly  subtle.   It  would   be
 interesting to understand their properties more precisely.

We end with some brief comments about other  avenues that deserve further exploration:

There are many other ways in which Eisenstein series and other automorphic forms could enter into the discussion of string theory scattering.  Clearly the maximal parabolic series of interest in this paper are but a small subset of the general series. Though we consider it unlikely for the examples here, cusp forms might also contribute to the solutions -- this would signify deep arithmetic complexity. Furthermore, a wider variety of constant terms, defined with respect to other maximal and non-maximal parabolic subgroups may well have a r{\^o}le to play in aspects of string theory and M-theory.

 The extension of the ideas in this paper to higher order terms in the derivative expansion raises very interesting new issues, specially since the next term, of order $\partial^8 \cR^4$, is no longer expected to be BPS so there is a strong possibility that its coefficient, $\calE^{(D)}_{(2,0)}$, gets contributions from all orders in string perturbation theory~\cite{Green:2010}.  Although there is some information about this function based on its M-theory limit in $D=9$~\cite{Green:2008bf}, this  is far from complete.  Beyond that, it is not at all clear how the discrete symmetry acts on the complete scattering amplitude.   Another clear challenge is the extension of these considerations to the coefficients in the low energy expansion of multiparticle scattering amplitudes.  Further afield are possible generalisations to amplitudes with non-maximal supersymmetry, which have lower-rank duality groups, or to scattering in curved space.

 \section*{Acknowledgements}
We  would  like to  thank  Ling  Bao,  Lisa Carbone,  Pierre  Cartier,
Freydoon Shahidi, Laurent Lafforgue, Erez Lapid, Peter Sarnak, and Wilfried Schmid for discussions.  S.M. acknowledges support from grant DMS-0901594, and
J.R. acknowledges support by MCYT  Research Grant No.  FPA 2007-66665 and Generalitat de
Catalunya under project 2009SGR502.  MBG is grateful for the support of a European Research Council Advanced Grant No. 247252.

\bigskip
\noindent {\bf Note}: As this paper was being written a
new appendix to~\cite{Pioline:2010kb} appeared making use of~\cite{GRS} and~\cite{Green:2010} to obtain
results related to limit (i) of this  paper.  These results appear to be related to those presented here upon making use of identities between different maximal parabolic Eisenstein series such as those in~\eqref{equiveisen}. As   explained at the end  section~\ref{sec:constantterm},  the complete analysis performed
  in this paper requires properties of Eisenstein series
  that cannot be obtained by the method used in~\cite{GRS}.
  Other recent papers~\cite{Lambert:2010pj} also cover related topics.

\vfill\eject

\appendix
\section{Tables}\label{sec:tables}
\subsection{Duality groups and maximal parabolic subgroups}~\begin{table}[h]
   \centering
   \begin{tabular}{||c|c|c|c|c||}
   \hline
 $D$ & rank & $E_{d+1}(\IR)$ & $K$&$E_{d+1}(\ZZ)$\\
 \hline
 10A&1&$ \IR^+$&1&1\\
 10B&1&$SL(2,\IR)$&$SO(2)$& $SL(2,\ZZ)$\\
 9&2&$SL(2,\IR)\times\IR^+$& $SO(2)$& $SL(2,\ZZ)$\\
 8&3&$SL(3,\IR)\times SL(2,\IR)$& $SO(3)\times SO(2)$& $SL(3,\ZZ)\times SL(2,\ZZ)$ \\
 7&4&$SL(5,\IR)$ & $SO(5)$& $SL(5,\ZZ)$\\
 6&5 &$ SO(5,5,\IR)$ & $SO(5)\times SO(5)$& $SO(5,5,\ZZ)$\\
 5&6&$ E_{6(6)}(\IR)$& $USp(8)/\ZZ_2$& $ E_{6(6)}(\ZZ)$\\
 4&7 &$ E_{7(7)}(\IR)$& $SU(8)/\ZZ_2$& $E_{7(7)}(\ZZ)$\\
 3&8 &$ E_{8(8)}(\IR)$& $Spin(16)/\ZZ_2$& $E_{8(8)}(\ZZ)$\\
 \hline
   \end{tabular}\vspace{1ex}
   \caption{ The moduli parameterise the coset $E_{d+1}(\IZ)\bs E_{d+1}(\IR)/K$.}
   \label{tab:Udual}
 \end{table}~\begin{table}[h]
\centering\begin{tabular}[]{||c|c|c|c||}
 \hline
 deleted~node&$E_8$&$E_7$&$E_6$\\
 \hline
right&$       E_7$&$       E_6$&$     SO(5,5)$\\
 left & $SO(7,7)$& $ SO(6,6)$&$SO(5,5)$\\
 upper&$ SL(8)$&$ SL(7)$&$ SL(6)$\\
 \hline
\hline
 deleted~node&$E_5=SO(5,5)$&$E_4= SL(5)$&$E_3=SL(3)\times SL(2)$\\
 \hline
right&$  SL(5)$&$ SL(3)\times SL(2)$&$SL(2)\times SL(2)$\\
 left & $ SO(4,4)$ & $ SO(3,3)$& $ SO(2,2)$\\
 upper&$ SL(5)$&$ SL(4)$&$ SL(3)$\\
 \hline
 \end{tabular}\vspace{1ex}
   \caption{
The parabolic subgroups associated with the simple roots $\alpha_{d+1}$, $\alpha_1$ and $\alpha_2$  of $E_{d+1}$.}
   \label{tab:Parabolics}
 \end{table}

\newpage
\subsection{Solutions        for       $\calE^{(D)}_{(0,0)}$       and
  $\calE^{(D)}_{(1,0)}$ for rank-$(10-D)$ duality groups}~\begin{table}[hp]
   \centering
   \begin{tabular}{||c|c|c||}
   \hline
 $G_{d}(\IZ)=E_{d+1}(\IZ)$ & $\calE^{(D)}_{(0,0)}$  & $\calE^{(D)}_{(1,0)}$\\[2ex]
 \hline
$ E_{8(8)}(\IZ)$&$\bE^{E_{8}}_{[1\,0^7];{3\over  2}}$  & $\frac12\,
 \bE^{E_{8}}_{[1\,0^7];{5\over 2}}$ \\[2ex]
$ E_{7(7)}(\IZ)$&$\bE^{E_{7}}_{[1\,0^6];{3\over  2}}$  & $\frac12\,
 \bE^{E_{7}}_{[1\,0^6];{5\over 2}}$ \\[2ex]
$  E_{6(6)}(\IZ)$& $\bE^{E_{6}}_{[1\,0^5];{3\over  2}}$  & $\frac12\,
 \bE^{E_{6}}_{[1\,0^5];{5\over 2}}$ \\[2ex]
$ SO(5,5,\IZ)$ & $\bE^{SO(5,5)}_{[10000];{3\over 2}}$  & $ \frac12\, \hbE^{SO(5,5)}_{[10000];{5\over 2}}+{4\over 45} \hbE^{SO(5,5)}_{[00001];{3}}$ \\[2ex]
$SL(5,\IZ)$   &   $\bE^{SL(5)}_{[1000];{3\over   2}}$   &   $\frac12\,
\hbE^{SL(5)}_{[1000];\frac52}+{\pi\over 30}\hbE^{SL(5)}_{[0010];\fiveh}$ \\[2ex]
$SL(3,\IZ)\times SL(2,\IZ)$& $\hbE^{SL(3)}_{[10];\threeh}  +2 \hbE_1(U) $  & $\frac12\,\bE^{SL(3)}_{[10];\fiveh}
 -4\, \bE^{SL(3)}_{[10];-\frac12}  \,\bE_2(U) $ \\[2ex]
$SL(2,\IZ)$& $\bE_{3\over2}(\Omega)\,\nu_1^{-{3\over7}} +4\zeta(2)\,\,\nu_1^{{4\over7}}$ & $\frac12\, \nu_1^{-{5\over 7}}\,\bE_{5\over2}(\Omega)+{2\zeta(2)\over15}\, \nu_1^{{9\over7}}$ \\[2ex]
$SL(2,\IZ)$& $\bE_{3\over 2}(\Omega)$  & $\frac12\, \bE_{\fiveh}(\Omega)$   \\[2ex]
 \hline
   \end{tabular}\vspace{1ex}
     \label{tab:Udual1}
     \caption{Solutions for  the coefficients in  $D=10-d$ dimensions.
       The variables $\nu_1$ and $\Omega$ parameterise the $GL(1)$ and
       $SL(2)/SO(2)$ factors in the $D=9$ moduli space.}
 \end{table}

\clearpage
\subsection{Constant terms for $\calE^{(D)}_{(0,0)}$ and $\cE^{(D)}_{(1,0)}$}~\begin{table}[htpb]
\centering
  \begin{tabular}{||c|c|c||}
  \hline
&  Decompactification limit &  String perturbation limit \\
$\calE^{(D)}_{(0,0)}$  & Constant term in $P_{\alpha_{d+1}}$ &   Constant term in $P_{\alpha_1}$\\
 & $r^2 =r_d/\ell_{D+1}$ & $r^{-4}=y_D$ \\
\hline
$\bE^{E_{8}}_{[1\,0^7];{3\over  2}}$ & ${3\zeta(5)\over  \pi}\ r^{20} + r^{12} \bE^{E_7}_{[1\,0^6];{3\over
    2}}$     &     $2\zeta(3)     r^{24}+r^{20}\    {3\over     2\pi}
\bE^{SO(7,7)}_{[1\,0^6];{5\over  2}}$  \\[2ex]
$\bE^{E_{7}}_{[1\,0^6];{3\over  2}}$ & ${4\zeta(4)\over \pi}\ r^{12}+ r^6  \bE^{E_6}_{[10000];{3\over 2}}$  & $2\zeta(3) r^{12}+r^8\ {2\over \pi} \bE^{SO(6,6)}_{[100000];2}$
 \\[2ex]
 $\bE^{E_{6}}_{[1\,0^5];{3\over  2}}$& $2\zeta(3) r^8+r^4  \bE^{SO(5,5)}_{[10000];{3\over 2}}$ & $2\zeta(3) r^8+ r^4   \bE^{SO(5,5)}_{[10000];\frac32}$ \\[2ex]
$\bE^{SO(5,5)}_{[10000];{3\over 2}}$&  $4\zeta(2) r^5+r^3\bE^{SL(5)}_{[1000];{3\over 2}}$  & $ 2\zeta(3) r^6+ 2 r^2\bE^{SO(4,4)}_{[1000];1} $ \\[2ex]
 $\bE^{SL(5)}_{[1000];{3\over 2}}$&$r^{12\over5}\,
(\hbE^{SL(3)}_{[10];\frac32}+2\hbE^{SL(2)}_1$&$2\zeta(3)r^{24\over5}+2\pi
r^{4\over5}\bE^{SO(3,3)}_{[100];\frac12}$\\
&$-8\,\pi\log r)$&\\[2ex]
 $\hbE^{SL(3)}_{[10];\threeh}  +2 \hbE^{SL(2)}_1 $&$r^2\,
(\nu_1^{-{3\over7}}\bE^{SL(2)}_{\frac32}+4\zeta(2)\nu_1^{4\over7})$&$2\zeta(3)      r^4+      2\bE^{SO(2,2)}_{[10];0}$\\
&$-{24\pi\over3}\log
r$&$-{8\pi\over3}\log
r$\\[2ex]
 $\nu_1^{-{3\over7}}\bE_{3\over2}(\Omega)+4\zeta(2)\nu_1^{{4\over7}}$&$4\zeta(2)r^{-{16\over7}}+r^{12\over7}\,\bE^{SL(2)}_{\frac32}$&$2\zeta(3)r^{24\over7}+r^{-{4\over7}}\bE^{SO(1,1)}_{-\frac12}$\\[2ex]
\hline
  \end{tabular}
\vspace{1ex}
 \caption{  The   constant  terms  of   $\calE^{(D)}_{(0,0)}$  in  the
   parabolic   subgroups   specified   by  limits~(i)   and~(ii)   for
   $D=3,\dots,9$.  The  scales
   in the logarithms have been  absorbed into the non-analytic part of
   the string amplitude.
\label{tab:Udual2}}
\end{table}

\begin{table}[h!]
\centering
  \begin{tabular}{||c|c||}
  \hline
& M-theory limit\\
$\calE^{(D)}_{(0,0)}$   & Constant term in $P_{\alpha_2}$\\
& $r^{(2+2d)/3} = \cV_{d+1}/\ell_{11}^{d+1}$  \\
\hline
$\bE^{E_{8}}_{[1\,0^7];{3\over  2}}$ &  $  4\zeta(2)  r^{32}+  r^{30}
\bE^{SL(8)}_{[1\,0^6];{3\over 2}} $ \\[2ex]
$\bE^{E_{7}}_{[1\,0^6];{3\over  2}}$
&      $ 4\zeta(2) r^{14}+ r^{12} \bE^{SL(7)}_{[1\,0^5];{3\over 2}} $   \\[2ex]
 $\bE^{E_{6}}_{[1\,0^5];{3\over  2}}$ & $4\zeta(2)\, r^8+r^6 \bE^{SL(6)}_{[10000];{3\over 2}}$\\[2ex]
$\bE^{SO(5,5)}_{[10000];{3\over 2}}$&
    $4\zeta(2)r^5+ r^3\bE^{SL(5)}_{[1000];\frac32}$  \\[2ex]
 $\bE^{SL(5)}_{[1000];{3\over 2}}$&$4\zeta(2)r^{16\over5}+r^{6\over5}\bE^{SL(4)}_{[100];\frac32}$\\
 $\hbE^{SL(3)}_{[10];\threeh}  +2 \hbE^{SL(2)}_1 $&$4\zeta(2) \, r^2+\hbE^{SL(3)}_{[10];\frac32}$\\
&$ -4\pi\log r$\\[2ex]
 $\nu_1^{-{3\over7}}\bE_{3\over2}(\Omega) +4\zeta(2)\nu_1^{{4\over7}}$&$4\zeta(2)
r^{8\over7} + r^{-{6\over7}}\bE^{SL(2)}_{\frac32}$\\[2ex]
\hline
  \end{tabular}\vspace{1ex}
  \caption{The   constant  terms   of  $\calE^{(D)}_{(0,0)}$   in  the
    parabolic     subgroups    specified     by     limit~(iii)    for
    $D=3,\dots,9$.  The scale
   in the logarithm has again been absorbed into the non-analytic part of
   the string amplitude.
\label{tab:Udual2b}}
\end{table}

\newpage
\begin{table}[h]
   \centering
   \begin{tabular}{|c||c||}
   \hline
   & Decompactification Limit\\
$\calE^{(D)}_{(1,0)}$ & Constant term in $P_{\alpha_{d+1}}$\\
& $ r^2 = r_d/\ell_{D+1}$. \\
 \hline
  $\frac12\,
 \bE^{E_{8}}_{[1\,0^7];{5\over 2}}$  & ${1\over 2} r^{20}  \bE^{E_7}_{[1\,0^6];{5\over 2}}+{\zeta(3)\over \pi }\, r^{24}
\bE^{E_7}_{[1\,0^6];{3\over 2}}+{7\zeta(9)\over 12\pi}\ r^{36}$
  \\
  $\frac12\,
 \bE^{E_{7}}_{[1\,0^6];{5\over 2}}$  & $ {1\over 2} r^{10}  \bE^{E_6}_{[1\,0^5];{5\over 2}} +
{\pi\over 3}\ r^{12}\bE^{E_6}_{[1\,0^5];{3\over 2}}   +{8\zeta(8)\over15\pi}\, r^{24}  $
\\
  $\frac12\,
 \bE^{E_{6}}_{[1\,0^5];{5\over 2}}$& $  r^{20\over 3} ( {1\over 2} \hbE^{SO(5,5)}_{[10000];{5\over 2}}+
{4\over       45}\,\hbE^{SO(5,5)}_{[00001];{3}})
+4\,r^{20\over3}\log r\, \bE^{SO(5,5)}_{[10000];\frac32}+
{\zeta(7)\over 6}\, r^{56\over 3} $ \\[2ex]
 $    \frac12\,   \hbE^{SO(5,5)}_{[10000];{5\over    2}}+{4\over   45}
 \hbE^{SO(5,5)}_{[00001];{3}}$& $r^5\,
 (\frac12\hbE^{SL(5)}_{[1000];\frac52}+{\pi\over30}\hbE^{SL(5)}_{[0010];\frac52}+2\pi^2\log
 r)-5
 r^3\log r\, \bE^{SL(5)}_{[1000];\frac32}+
{8\zeta(6)\over45}   \,   r^{15}
$ \\[2ex]
 \hline
    \end{tabular}\vspace{1ex}
   \caption{The constant terms  of  $\calE^{(D)}_{(1,0)}$
in  the  parabolic  subgroups  specified by  limit~(i)  in  dimensions
$D=3,4,5,6$.  The scale
   in the logarithm has again been absorbed into the non-analytic part of
   the string amplitude. 
   \label{tab:Udual3} }
 \end{table}~\begin{table}[h]
   \centering
   \begin{tabular}{|c|c||}
   \hline
     & String perturbation limit\\
   $\calE^{(D)}_{(1,0)}$   & Constant term in $P_{\alpha_1}$  \\
   &$r^{-4}=y_D$ \\
 \hline
  $\frac12\,
 \bE^{E_{8}}_{[1\,0^7];{5\over 2}}$           &          $           r^{40}\zeta(5)+{7\over 24\pi}\, r^{36} \bE^{SO(7,7)}_{[1\,0^6];{9\over 2}}+{2\over          3}
 r^{32}\bE^{SO(7,7)}_{[0^5\,1\,0];{2}} $ \\[2ex]
   $\frac12\,
 \bE^{E_{7}}_{[1\,0^6];{5\over 2}}$
& $  r^{20}\zeta(5)+{4\over 15\pi}\, r^{16}
\bE^{SO(6,6)}_{[1\,0^5];4} +{2\over 3} r^{12}\bE^{SO(6,6)}_{[0^4\,1\,0];{2}}$\\[2ex]
   $\frac12\,
 \bE^{E_{6}}_{[1\,0^5];{5\over 2}}$ & $  r^{40\over 3}\zeta(5)+{1\over 12}\, r^{28\over 3}
\bE^{SO(5,5)}_{[10000];{7\over 2}}+{2\over 3} r^{16\over 3}
\bE^{SO(5,5)}_{[00010];{2}} $ \\[2ex]
 $    \frac12\,   \hbE^{SO(5,5)}_{[10000];{5\over    2}}+{4\over   45}
 \hbE^{SO(5,5)}_{[00001];{3}}$&
 $r^{10}\zeta(5)+2\zeta(3)\,r^6\,\partial_s\bE^{SO(4,4)}_{[0001];0}-
 4\zeta(3)r^6\,\log r$\\
&$+ {2\over3} r^2\,
(\hbE^{SO(4,4)}_{[1000];2}+\hbE^{SO(4,4)}_{[0001];2})-4
\bE^{SO(4,4)}_{[1000];1}\, r^2\log r$\\
 \hline
    \end{tabular}\vspace{1ex}
   \caption{\label{tab:Udual4}
The constant terms of
$\calE^{(D)}_{(1,0)}$  in the parabolic  subgroups specified  by limit
(ii) in dimensions  $D=3,4,5,6$. The scale in the  logarithm has again
been absorbed into the non-analytic part of the string amplitude
}
 \end{table}~\begin{table}[h]
   \centering
   \begin{tabular}{|c||c||}
   \hline& M-theory limit\\
$\calE^{(D)}_{(1,0)}$&Constant term in $P_{\alpha_2}$  \\
& $r^{(2+2d)/3} =\cV_{d+1}/\ell^{d+1}_{11}$\\
 \hline
  $\frac12\,
 \bE^{E_{8}}_{[1\,0^7];{5\over 2}}$& $\frac12r^{50}\bE^{SL(8)}_{[1\,0^6];\frac52}+{2\zeta(3)\over\pi^2} r^{48}\bE^{SL(8)}_{[0\,1\,0^5];2}-8\zeta(4)\,r^{54}\bE^{SL(8)}_{[1\,0^6];-\frac12}$ \\[2ex]
  $\frac12\,
 \bE^{E_{7}}_{[1\,0^6];{5\over 2}}$&$\frac12 r^{20}\bE^{SL(7)}_{[1\,0^5];\frac52}+{2\zeta(3)\over\pi^2}r^{18}\bE^{SL(7)}_{[01\,0^5];2}-8\zeta(4)\,r^{24}\bE^{SL(7)}_{[1\,0^5];-\frac12}$ \\[2ex]
  $\frac12\,
 \bE^{E_{6}}_{[1\,0^5];{5\over 2}}$ &$\frac12 r^{10}\bE^{SL(6)}_{[1\,0^4];\frac52}+{2\zeta(3)\over\pi^2}r^{8}\bE^{SL(6)}_{[0\,1\,0^3];2}-8\zeta(4)\,r^{14}\bE^{SL(6)}_{[1\,0^4];-\frac12}$\\[2ex]
  $   \frac12\,    \hbE^{SO(5,5)}_{[10000];{5\over   2}}+{4\over   45}
  \hbE^{SO(5,5)}_{[00001];{3}}$   & $ r^5(\frac12 \hbE_{[10000];\frac52} -2\pi^2 \log r)-8\zeta(4)\,r^9  \bE^{SL(5)}_{[1000];-\frac12}
+$\\
&$
+r^3({\zeta(3)\over3\zeta(2)}\hbE^{SL(5)}_{[0100];2} - \partial_s\bE^{SL(5)}_{[1000];\frac32}     -
{\zeta(3)\over\zeta(2)} \textrm{Res}_{s=2}\bE^{SL(5)}_{[0100];s} \,\log r)
$ \\[2ex]
 \hline
    \end{tabular}\vspace{1ex}
   \caption{The constant terms  of  $\calE^{(D)}_{(1,0)}$
in  the parabolic  subgroups specified  by limit  (iii)  in dimensions
$D=3,4,5,6$.  
  The  scales
   in the logarithms have again been absorbed into the non-analytic part of
   the string amplitude. 
\label{tab:Udual5} }
 \end{table}

\end{document}